\documentclass[a4paper,11pt]{article}
\usepackage{jheppub} % for details on the use of the package, please see the JINST-author-manual
\usepackage{lineno}
\usepackage{mathtools}
\usepackage{braket}

\def\bal#1\eal{\begin{align}#1\end{align}}

\preprint{TU-1269}
\title{Decay of a scalar condensate in two different approaches}

\author{
{\large Ayuki Kamada$^{1}$, Kodai Sakurai$^{1,2,3}$}
\\*[20pt]
{\it \normalsize 
$^1$Institute of Theoretical Physics, Faculty of Physics, University of Warsaw, \\
ul.~Pasteura 5, PL-02-093 Warsaw, Poland \\[5pt]
$^2$Department of Physics, Tohoku University, 
Sendai, Miyagi 980-8578, Japan \\[5pt]
$^3$National Institute of Technology, Tsuruoka College, Tsuruoka, Yamagata 997-0842, Japan
} \\*[5pt]
}

\emailAdd{akamada@fuw.edu.pl, kodai.sakurai@tsuruoka-nct.ac.jp}

\abstract{
Decay of a scalar condensate via interactions with (quasi-)particles is of interest to many fields in physics, including cosmology.
In cosmology, the decay of an inflaton condensate leads to the production of daughter particles and reheating of the Universe.
In computing the decay rate, two quantum field theoretic approaches can be found in the literature: one is based on parametric resonance of mode functions of the daughter particle; another is based on the $S$-matrix of a coherent state and Feynman-diagrammatic perturbation theory.
We modify the latter from the previous literature in a way that manifests what we are computing and does not include unwanted Feynman diagrams.
We notice the equivalence of these two approaches and demonstrate it by explicitly computing the decay rate at lower orders in the double expansion of the amplitude of coherent oscillation (or narrow resonance) and velocity of the daughter particle.
}

\begin{document}
\maketitle
\flushbottom

\section{Introduction}\label{sec:intro}
A scalar condensate appears in various fields of physics.
A notable example is inflaton in cosmology which drives an inflation period and then starts to oscillate in the very early Universe~\cite{Kolb:1990vq}.
A scalar condensate is made of a macroscopic number of scalar particles (and thus not necessarily a number eigenstate) and described by a field (or macroscopic wave function) which follows a classical field equation (or Gross-Pitaevskii equation).
A scalar condensate dissipates energy through interaction with other particles including the production of light particles (or excitations).
Energy dissipation (or decay) of an inflaton condensate leads to reheating of the Universe and thus often has phenomenological or observational consequences (for example, reheating temperature should be higher than a few MeV for successful big-bang nucleosynthesis)~\cite{Hasegawa:2019jsa}.

Motivated by this, we consider the decay of a scalar condensate oscillating coherently over a space without any ambient gas.
The most minimal approach is to consider the decay of a scalar particle into daughter particles.
This approach misses two effects.
The first is that since a scalar condensate is made of a macroscopic number of scalar particles, not only decay but also $N(\ge2)$ processes ($N=2$ is annihilation) can produce daughter particles.
The second is that a scalar-field value alters the microscopic properties of daughter particles such as mass.
These effects are expected to be important especially for a large field value.
It is indeed known that these two effects alter energy dissipation of a scalar condensate drastically, especially when a field value is large~{\cite{Kofman:1994rk,Dolgov:1989us,Traschen:1990sw}}.
Time-changing mass of daughter particles leads to an exponentially growing mode function called parametric resonance.
Since energy dissipation proceeds more quickly than the decay of scalar particles, it is called preheating.

Exponential growth of the mode function needs a seed, and a natural candidate is quantum (daughter-particle number) fluctuation.
Here one needs to take into account some quantum field theory.
Actually, Ref.~\cite{Yoshimura:1995gc} formulates daughter-particle production or energy dissipation of a scalar condensate as a vacuum-to-vacuum transition of quantized daughter particles in the presence of a scalar condensate.
Meanwhile, Ref.~\cite{Matsumoto:2007rd} formulates decay of a scalar condensate in terms of $S$-matrix in quantum field theory, by postulating that a scalar condensate is described by a coherent state.
This allows one to compute the decay of a scalar condensate in a Feynman diagrammatic way similar to ordinary perturbation theory.
On the other hand, it is not clear how the $S$-matrix is computed for a coherent state and interpreted as a decay rate.
The relation to the parametric-resonance approach~\cite{Yoshimura:1995gc} is also not clear.
Ref.~\cite{Matsumoto:2007rd} demonstrates that the Feynman-diagrammatic approach gives the same result as the parametric-resonance approach, but only in the limit where a daughter particle mass is close to the half of a scalar mass.

In this paper, we establish the relation between the Feynman-diagrammatic approach and the parametric-resonance approach.
A key notion is that the sum of ring diagrams with all possible insertions of a scalar field provides an effective action of a scalar field, namely, vacuum-to-vacuum transition.
Therefore, these two approaches indeed compute the same quantity but in different ways (one is solving the Schrodinger equation; the other is performing a path integral).
Not only they look different, but the quantities one can easily compute in the two approaches are also different.
After clarifying this difference, we demonstrate that these two computations coincide with each other by explicitly computing the double expansion of the decay rate of a scalar condensate in terms of the oscillation amplitude and velocity of the daughter particle.

Meanwhile, in both approaches, a scalar condensate is regarded as a background.
Though this is a good approximation in an early stage of preheating, one needs to take into account back reactions to a scalar condensate in a late stage.
It is known that not only energy dissipation but also large occupation number of daughter particles alter the evolution of a scalar condensate~\cite{Kofman:1997yn}.
The most first-principle approach would be adopting non-equilibrium quantum field theory~\cite{Berges:2004yj,Calzetta:2008iqa}~{(also see Refs.~\cite{Morikawa:1986rp,Boyanovsky:1994me,Greiner:1996dx,Yokoyama:2004pf,Berera:2008ar,Mukaida:2013xxa,Wang:2022mvv,Ai:2023ahr})}. 
%~\footnote{{We note that this}}.
A popular approach is performing a classical lattice simulation with quantum fluctuation being a seed~\cite{Felder:2000hq}.
It is beyond the scope of this paper to study such a late stage of preheating.

This paper is organized as follows.
In the next section, we review the parametric-resonance approach (Sec.~\ref{sec:review parametric}) and Feynman-diagrammatic approach (Sec.~\ref{sec:review diagrammatic}).
We also modify the latter in a way that manifests what we are computing and does not include the unwanted Feynman diagrams.
After noticing the two approaches actually compute the same quantity (Sec.~\ref{sec:comparison}), we introduce the double expansion of the decay rate of a scalar condensate to manifest the apparent difference between the two methods.
We compute the decay rate at lower orders of the expansion in the parametric-resonance approach (Sec.~\ref{sec:parametric}) and Feynman-diagrammatic approach (Sec.~\ref{sec:diagrammatic}).
After checking they give the same result (Sec.~\ref{sec:equality}), we provide concluding remarks (Sec.~\ref{sec:conclusion}).
We devote appendix~\ref{sec:apparent divergence} to discussing an apparent divergence appearing in the Feynman-diagrammatic computation, and appendix~\ref{sec:oscillation average} to discussing yet another method in computing decay rate of a scalar condensate.

\section{Review of two methods}

By following Refs.~\cite{Yoshimura:1995gc, Matsumoto:2007rd}, we consider a simple model whose Lagrangian is given by
\bal
{\cal L} \supset \frac{1}{2} (\partial \varphi)^{2} + \frac{1}{2} (\partial \chi)^{2} - \frac{1}{2} m_{\varphi}^{2} \varphi^{2} - \frac{1}{2} m_{\chi}^{2} \chi^{2} + \frac{1}{2} \mu \varphi \chi^{2} \;.
\eal
A scalar condensate is made of a real scalar $\varphi$, and a daughter particle is a real scalar $\chi$.
Energy dissipation proceeds through $\varphi \chi^2$ interaction.
We assume that other (self-)interactions have a tiny coupling and only play a subdominant role.
Equations of motion are
\bal
\partial^{2} \varphi + m_{\varphi}^{2} \varphi - \frac{1}{2} \mu \chi^{2} = 0 \;. \\
\partial^{2} \chi + m_{\chi}^{2} \chi - \mu \varphi \chi = 0 \;.
\eal
At the level of background, we assume $\chi = 0$ and spatial coherence of $\varphi$, and thus the solution is
\bal
\varphi = \varphi_{+} + \varphi_{-} \;, \quad
\varphi_{\pm} = \frac{1}{2} A_{\varphi} e^{\mp i m_{\varphi} t} \;.
\eal
In the presence of $\varphi$ background, the equation of motion for $\chi$ is given by
\bal
\partial^{2} \chi + m^{2}_{\chi {\rm eff}} (t) \chi = 0 \;, \quad m^{2}_{\chi {\rm eff}} (t) = m_{\chi}^{2} - \mu A_{\varphi} \cos(m_{\varphi} t) \;.
\eal
The effective mass is changing with time. 
In particular, when $\mu A_{\varphi} > m_{\chi}^{2}$, it becomes tachyonic in a certain time interval.
Hereafter, we do not consider a fluctuation of $\varphi$ {unless stated}.

\subsection{Parametric-resonance approach} \label{sec:review parametric}

In the presence of a $\varphi$ background, $\chi$ can be regarded as a free field but with a time-changing effective mass.
Therefore, we can quantize $\chi$ in an ordinary way with the same-time commutator (0 for the other commutators) and Hamiltonian:
\bal
[{\hat \chi}(\vec{x}, t), {\hat \pi}_{\chi}(\vec{x}', t)] = i \delta^{3} (\vec{x} - \vec{x}') \;, \\
%\quad {\hat \pi}_{\chi}(\vec{x}, t) = \frac{\partial}{\partial t} {\hat \chi}(\vec{x}, t) \;, \\
{\hat H} (t) = \int d^{3}x \left[ \frac{1}{2} {\hat \pi}_{\chi}^{2} + \frac{1}{2} (\nabla {\hat \chi})^{2} + \frac{1}{2} m^{2}_{\chi {\rm eff}} (t) {\hat \chi}^{2} \right] \;.
\eal
Because of the space-translation/rotation symmetry, it is more convenient to consider a Fourier space rather than a real space (same for ${\hat \pi}_{\chi}$):
\bal
{\hat \chi}(\vec{x}, t) = \int \frac{d^{3}k}{(2 \pi)^{3}} \frac{1}{\sqrt{2}}\left( {\hat \chi}_{R\vec{k}} (t) + i {\hat \chi}_{I\vec{k}} (t) \right) e^{i \vec{k} \cdot \vec{x}} \;, \\
{\hat \chi}_{R\vec{k}} (t) = {\hat \chi}_{R\vec{k}}^{\dagger} (t) = {\hat \chi}_{R-\vec{k}} (t) = {\hat \chi}_{R -\vec{k}}^{\dagger} (t) \;, \\
{\hat \chi}_{I\vec{k}} (t) = {\hat \chi}_{I\vec{k}}^{\dagger} (t) = - {\hat \chi}_{I-\vec{k}} (t) = - {\hat \chi}_{I -\vec{k}}^{\dagger} (t) \;.
\eal
Here we define ${\hat \chi}_{R\vec{k}} (t)$ and ${\hat \chi}_{I\vec{k}} (t)$ so that they are Hermitian, and use the fact that ${\hat \chi}(\vec{x}, t)$ is Hermitian.
One can rewrite the same-time commutator and Hamiltonian as
\bal
[{\hat \chi}_{R\vec{k}}(t), {\hat \pi}_{\chi R \vec{k}'}(t)] = [{\hat \chi}_{I\vec{k}}(t), {\hat \pi}_{\chi I \vec{k}'}(t)] = i (2 \pi)^{3} \delta^{3} (\vec{k} - \vec{k}') \;, \\
%{\hat \pi}_{R \chi \vec{k}}(t) = \frac{d}{d t} {\hat \chi}_{R \vec{k}}(t) \,, \quad {\hat \pi}_{I \chi \vec{k}}(t) = \frac{d}{d t} {\hat \chi}_{I \vec{k}}(t) \;,\\
{\hat H} (t) = \int \frac{d^{3}k}{(2 \pi)^{3}} \left[ \frac{1}{4} \left( {\hat \pi}_{R\chi \vec{k}}^{2} + {\hat \pi}_{I\chi \vec{k}}^{2} \right) + \frac{1}{4} E^{2}_{\chi k} (t) \left( {\hat \chi}_{R \vec{k}}^{2} + {\hat \chi}_{I \vec{k}}^{2} \right) \right] \;,
\eal
where $E^{2}_{\chi k} (t) = k^{2} + m^{2}_{\chi {\rm eff}} (t)$.
Because ${\hat \chi}_{R\vec{k}} (t) = {\hat \chi}_{R-\vec{k}} (t)$ and ${\hat \chi}_{I \vec{k}} (t) = - {\hat \chi}_{I-\vec{k}} (t)$, modes with ${\vec k}$ and $-{\vec k}$ are not independent. 
On the other hand, since the Hamiltonian is symmetric under the exchange of ${\hat \chi}_{R\vec{k}} (t)$ and ${\hat \chi}_{I\vec{k}} (t)$, this quantum system is equivalent to
\bal
[{\hat q}_{\vec{k}}(t), {\hat p}_{\vec{k}'}(t)] = i (2 \pi)^{3} \delta^{3} (\vec{k} - \vec{k}') \;, \quad \quad
%\quad {\hat p}_{\vec{k}}(t) = \frac{d}{d t} {\hat q}_{\vec{k}}(t) \;, \\
{\hat H} (t) = \int \frac{d^{3}k}{(2 \pi)^{3}} \left[ \frac{1}{2} {\hat p}_{\vec{k}}^{2} + \frac{1}{2} E^{2}_{\chi k} (t) {\hat q}_{\vec{k}}^{2} \right] \;,
\eal
where modes with different ${\vec k}$ are independent.

Let us consider a periodic boundary condition inside a cubic with a length $L$ (we take $L \to \infty$ in the end), so that ${\vec k}$ only takes discrete values.
Then, the quantum system is given by (under the rescaling of ${\hat q}_{\vec{k}}(t) \to {\hat q}_{\vec{k}}(t) L^{3/2}$ and ${\hat p}_{\vec{k}}(t) \to {\hat p}_{\vec{k}}(t) L^{3/2}$)
\bal
[{\hat q}_{\vec{k}}(t), {\hat p}_{\vec{k}'}(t)] = i \delta_{\vec{k},  \vec{k}'} \;, \quad \quad
%\quad {\hat p}_{\vec{k}}(t) = \frac{d}{d t} {\hat q}_{\vec{k}}(t), \\
{\hat H} (t) = \sum_{\vec{k}} \left[ \frac{1}{2} {\hat p}_{\vec{k}}^{2} + \frac{1}{2} E^{2}_{\chi k} (t) {\hat q}_{\vec{k}}^{2} \right] \;.
\eal
Since modes with different ${\vec k}$ are independent harmonic oscillators, the total wave function (wave functional) is given by a product of a wave function for each ${\vec k}$:
\bal
\Psi(\{q_{\vec{k}}\}, t) = \prod_{\vec{k}} \psi(q_{\vec{k}}, t) \;.
\eal
The wave function for each $k$ (we assume isotropy) satisfies
\bal
i \frac{\partial}{\partial t} \psi(q_{k}, t) = - \frac{1}{2} \frac{\partial^{2}}{\partial q_{k}^{2}} \psi(q_{k}, t) + \frac{1}{2} E^{2}_{\chi k} (t) q_{k}^{2} \psi(q_{k}, t) \;.
\eal
It admits a solution in the Gaussian form of\footnote{It may be worth noting that mass dimensions of $q_{k}$, $p_{k}$,  $\psi(q_{k}, t)$ and $u_{k} (t)$ are $-1/2$, $1/2$, $1/4$ and $-1/2$ respectively.}
\bal
\psi(q_{k}, t) = \frac{1}{u_{k}(t)^{1/2}} \exp \left[\frac{i}{2} \left( \frac{1}{u_{k}(t)} \frac{d}{dt} u_{k}(t) \right) q_{k}^{2} \right] \;,
\eal
where $u_{k}$ is a mode function satisfying
\bal
\frac{d^{2}}{dt^{2}} u_{k} + E^{2}_{\chi k} (t) u_{k} = 0 \;.
\eal

Since we are interested in vacuum evolution, we take
\bal
u_{k} (0) = \left( \frac{E_{\chi k} (0)}{\pi} \right)^{-1/2} \;, \quad \quad \frac{1}{u_{k}(0)} \frac{d}{dt} u_{k}(0) = i E_{\chi k} (0) 
\eal
as an initial condition so that the initial state is given by vacuum (or ground state):
\bal
\psi(q_{k}, 0) = \left( \frac{E_{\chi k} (0)}{\pi} \right)^{1/4} \exp \left[ - \frac{1}{2} E_{\chi k} (0) q_{k}^{2} \right] \;.
\eal
{Then, vacuum-to-vacuum amplitude (with time interval T) is computed as
\bal
_{\rm out}\langle 0 | 0 \rangle_{\rm in} &= \prod_{\vec{k}} \int dq_{k} \psi(q_{k}, 0) \psi(q_{k}, T) \notag \\
&= \prod_{\vec{k}} \left( 4 \pi E_{\chi k} (0) \right)^{1/4} \left[ E_{\chi k} (0) u_{k}(T) -  i \frac{d}{dt} u_{k}(T) \right]^{-1/2} \;.
\eal
By noting the Wronskian
\bal
u^{*}_{k}(t) \frac{d}{dt} u_{k}(t) - u_{k}(t) \frac{d}{dt} u^{*}_{k}(t) = 2 i \pi
\eal
with the initial condition above, one can rewrite its magnitude as
\bal
\label{eq:vac-to-vac}
|_{\rm out}\langle 0 | 0 \rangle_{\rm in}| &= \prod_{\vec{k}} \left( 4 \pi E_{\chi k} (0) \right)^{1/4} \left( E_{\chi k}^{2} (0) |u_{k}(T)|^{2} + 2 \pi E_{\chi k} (0) + \left|\frac{d}{dt} u_{k}(T) \right|^{2} \right)^{-1/4} \notag \\
&= \prod_{\vec{k}} \left( 1 + f_{\chi k} (T) \right)^{-1/4} \;.
\eal
Here 
\bal
\label{eq:phase_space}
f_{\chi k} (T) = \frac{1}{4 \pi E_{\chi k} (0)} \left( E_{\chi k}^{2} (0) |u_{k}(T)|^{2} + \left|\frac{d}{dt} u_{k}(T) \right|^{2} \right) - \frac{1}{2}
\eal
can be interpreted as the phase-space distribution function of produced $\chi$ particles~\cite{Yoshimura:1995gc} (see also Ref.~\cite{Kofman:1997yn} but note a different normalization of the mode function; also note that this interpretation is valid only when $T$ is a multiple of the oscillation period of the background field).
When some modes admit exponential growth $u_{k} \propto e^{\lambda_{k} m_{\varphi} t / 2}$,
\bal
\label{eq:Gamma_wrt_k}
_{\rm out}\langle 0 | 0 \rangle_{\rm in} \propto e^{- \Gamma L^{3}T / 2} \;, \quad
\Gamma = \frac{1}{L^{3}} \sum_{\vec{k}} \frac{m_{\varphi}}{2} \lambda_{k} = \frac{m_{\varphi}}{2} \int \frac{d^{3} k}{(2 \pi)^{3}} \lambda_{k} \;,
\eal
at large $T$.}
Here, a summation and integral over $k$ are limited for exponentially growing modes, and $\Gamma$ can be interpreted as a volumetric decay rate
of a scalar condensate.

The remaining task is to find an exponentially growing mode $u_{k}$.
We note that the equation of motion can be written in the following dimensionless form:
\bal
\label{eq:Mathieu}
\frac{d^2u_{ k}(z) }{dz^2}+\Big\{h-\theta 
{\rm exp}(-2iz) -\theta  {\rm exp}(+2iz)\Big\} u(z)=0
\eal
with the dimensionless time $z$ being $z= m_\varphi t/2$, where 
the other dimensionless energy $h$ and amplitude $\theta$ are given by
\bal
\label{eq:Matheiu_eq}
h&=4\frac{k^2+m_\chi^2}{m_\varphi^2}\;,\quad \quad
\theta =2\frac{\mu A_\varphi}{m^2_\varphi}\;. 
\eal
This is known as Mathieu equation, which admits a solution in the form of 
\bal
\label{eq:special solution}
u = e^{\mu z} P(z)
\eal
where $P(z)$ is a periodic function $P(z)=P(z+\pi)$ and $\mu$ is a complex-valued Floquet exponent.
It is known where $\mu$ takes a real value (instability chart) in the $(h, \theta)$ plane (see Ref.~\cite{Kofman:1994rk} for example, noting $h \leftrightarrow A$ and $\theta \leftrightarrow  q$).
For $q \ll 1$, instability chart is localized around certain values of $h = N^{2}$ with $N = 1, 2 \dots$ (narrow resonance).
This can be interpreted as $N$-body process: $N\varphi \to 2 \chi$.
Motivated by this we denote $N$ by $n_{\varphi}$ below.
For $q \gg 1$, instability chart is distributed over a broad range of $h$ (broad resonance).
Though broad resonance has a more significant impact for preheating, we restrict our discussion within the narrow-resonance regime, since the former is not tractable diagrammatically as we will see soon.

\subsection{Feynman-diagrammatic approach} \label{sec:review diagrammatic}
Ref.~\cite{Matsumoto:2007rd} formulates the decay of scalar condensate by using a coherent state and $S$-matrix.
A coherent state is an eigenstate of the annihilation operator.
The Fourier expansion of $\hat \varphi$ operator is given by
\bal
{\hat \varphi}(x) = \int \frac{d^{3}k}{(2 \pi)^{3}} \frac{1}{\sqrt{2 E_{\varphi k}}}\left( {\hat a}_{\vec{k}} e^{- i k \cdot x} +  {\hat a}^{\dagger}_{\vec{k}} e^{i k \cdot x} \right) \;, \\
[{\hat a}_{\vec{k}}, {\hat a}^{\dagger}_{\vec{k}'}] = (2 \pi)^{3} \delta^{3} (\vec{k} - \vec{k}') \;.
\eal
Let us again consider a periodic boundary condition inside a cubic with a length $L$ and then,
\bal
{\hat \varphi}(x) = \sum_{\vec{k}} \frac{1}{\sqrt{2 E_{\varphi k} L^{3}}}\left( {\hat a}_{\vec{k}} e^{- i k \cdot x} +  {\hat a}^{\dagger}_{\vec{k}} e^{i k \cdot x} \right) \;, \\
[{\hat a}_{\vec{k}}, {\hat a}^{\dagger}_{\vec{k}'}] = \delta_{\vec{k} ,  \vec{k}'} \;.
\eal
A coherent state is given by
\bal
| \varphi \rangle = \prod_{\vec{k}} e^{\alpha_{\vec{k}} {\hat a}^{\dagger}_{\vec{k}} - \alpha^{*}_{\vec{k}} {\hat a}_{\vec{k}}} | 0 \rangle
\eal
where
\bal
\varphi = \sum_{\vec{k}} \frac{1}{\sqrt{2 E_{\varphi k} L^{3}}}\left( \alpha_{\vec{k}} e^{- i k \cdot x} + \alpha^{*}_{\vec{k}} e^{i k \cdot x} \right) \;.
\eal
Coherent states satisfy
\bal
{\hat a}_{\vec{k}} | \varphi \rangle = \alpha_{\vec{k}} | \varphi \rangle \;, \quad \quad \langle \varphi' | \varphi \rangle = \prod_{\vec{k}} e^{- |\alpha_{\vec{k}}|^{2} - |\alpha'_{\vec{k}}|^{2} + 2 \alpha'^{*}_{\vec{k}} \alpha_{\vec{k}}} \;, \quad \quad \prod_{\vec{k}} \int \frac{d^{2} \alpha_{\vec{k}}}{\pi} | \varphi \rangle \langle \varphi | = {\hat I} \;.
\eal
and notably normal-ordered products can be replaced by field values:
\bal
\langle \varphi | N\{{\hat \varphi}(x_{1}) \dots\} | \varphi \rangle = \varphi (x_{1}) \dots \;.
\eal
Therefore, the Wick theorem for time-ordered produces leads to (see, e.g., Ref.~\cite{Peskin:1995ev})
\bal
\label{eq:Wick theorem}
\langle \varphi | T\{{\hat \varphi}(x_{1}) \dots\} | \varphi \rangle = \varphi (x_{1}) \dots + \text{all possible contractions and replacements} \;.
\eal
Here ``all possible contractions and replacements'' means that there will be one term for each possible way of contracting fields, where the uncontracted fields are replaced by field values. {Each contraction gives a propagator including quantum fluctuation of $\varphi$.}

Choosing
\bal
\alpha_{0} = \sqrt{2 m _{\varphi} L^{3}} \frac{1}{2} A_{\varphi} \;,
\eal
and the others $0$ leads to a scalar condensate oscillating coherently over a space.

$S$-matrix element (or $\alpha \to \beta$ transition amplitude) is given by an inner product between in and out states (belonging to the Hilbert space of full Hamiltonian ${\hat H}$) in Heisenberg picture (see, e.g., Ref.~\cite{Weinberg:1995mt}):
\bal
S_{\beta \alpha} =\; _{\rm out}\langle \beta | \alpha \rangle_{\rm in} \,.
\eal
The same matrix element is given by the free states (belonging to the Hilbert space of free Hamiltonian ${\hat H}_{0}$) with ${\hat S}$ operator (note that the definition is different in some literature) in the interaction picture:
\bal
S_{\beta \alpha} = \langle \beta |{\hat S}| \alpha \rangle \;, \quad \quad
{\hat S} = T\left\{\exp \left(- i \int dt {\hat V}(t)  \right) \right\}
\eal
where ${\hat H} = {\hat H}_{0} + {\hat V}$.
Note that ${\hat V}$ should include not only ``interaction'' terms of a Lagrangian but also field/mass/vacuum-energy counter terms in on-shell renormalization, so that the free Hamiltonian has the same energy spectrum as the full Hamiltonian.

By subtracting the trivial transition, one would define ${\hat S} = {\hat I} + i {\hat T}$.
Since ${\hat S}$ operator is Unitary, by considering the matrix element with $\beta = \alpha$, one can find the optical theorem:
\bal
{\rm Im} T(\alpha \to \alpha) = \frac{1}{2} |T(\alpha \to \text{all})|^{2} \;.
\eal
Here, note that ``all'' includes not only decay but also (uninteresting) self-scattering.
For example, let us consider $4 \to 4$ forward scattering in $\lambda \phi^4$ theory.
${\rm Im} T$ includes two $2 \to 2$ scatterings between 2 out of 4 particles, and corresponding $|T|^{2}$ is one $2 \to 2$ scattering with the other two particles unscattered.
The latter is not of interest as a 4-body process (we are often interested in connected $S$-matrix), but inevitable because of a cluster decomposition of $S$-matrix. (This is often not a problem in a 1 or 2-body process.)
Another example is a vacuum-to-vacuum transition in the presence of a background field.
The in-state vacuum and out-state vacuum are the same (vacuum is unique in the Hilbert space of full Hamiltonian), but up to a phase.
This phase is related to vacuum energy and thus 0 by the vacuum-energy counter term in the absence of background fields, but non-zero in their presence (see, e.g., Ref.~\cite{Weinberg:1996kr}).

Actually, Ref.~\cite{Matsumoto:2007rd} selects only relevant diagrams when computing 
decay rate via ${\rm Im} T$.
Instead, we suggest computing $S$-matrix itself and find a behavior of 
\bal
\label{eq:Gamma_wrt_S}
\langle \varphi |{\hat S}| \varphi \rangle \propto e^{- \Gamma L^{3} T /2}
\eal
with time interval $T$.
$\Gamma$ is again interpreted as a volumetric decay rate of a scalar condensate.
Thanks to the above property of coherent state~\eqref{eq:Wick theorem}, $\Gamma L^{3} T /2$ is simply given by the imaginary part of an effective action after integrating out fluctuations on top of the background. (Precisely speaking, one needs to subtract the free action of the background, but it does not change the imaginary part.)
The effective action is given by the sum of connected diagrams with external lines of the background and internal lines of fluctuations.
Its imaginary part can be computed by applying ordinary cutting rules on internal lines.
This procedure is similar to finding a 1PI effective action, but remember that the latter takes into account only connected 1PI diagrams.

In our toy model, we can consider an action of $S[\varphi + \delta \varphi, \chi]$ and compute connected diagrams with $\delta \varphi$ and $\chi$ internal lines. {Though we take into account $\delta \varphi$ above (remember that contraction $\varphi$ is identical to contraction of $\delta \varphi$),} in the rest of this paper, following Ref.~\cite{Matsumoto:2007rd}, we do not consider $\delta \varphi$.
Then, $\langle \varphi |{\hat S}| \varphi \rangle$, with ${\hat V} = \mu \varphi {\hat \chi}^{2} /2$ reduces to vacuum-to-vacuum transition of $\chi$ in the presence of $\varphi$ background (where effective action is given by the sum of 1-loop diagrams of $\chi$ with external lines of $\varphi$; see Fig.~1 of Ref.~\cite{Matsumoto:2007rd}).
This is exactly the same as what is considered in the previous section: $_{\rm out}\langle 0 | 0 \rangle_{\rm in}$ (where $|0\rangle$ is the vacuum of $\chi$ in the presence of $\varphi$ background).

{Before concluding this section, we comment on possible effects of $\delta \varphi$. It leads to production of on-shell $\delta \varphi$'s and quantum correction from virtual $\delta \varphi$'s. The former does not change production of $\chi$'s as far as we treat $\varphi$ as a classical background. Furthermore, production of $\delta \varphi$'s is loop-suppressed compared to production of $\chi$'s and thus subdominant in the perturbative regime $\mu \ll m_{\varphi}$. In this regime, quantum correction from virtual $\delta \varphi$'s is subdominant.}

\subsection{Comparison of the two methods}\label{sec:comparison}

In the previous sections, we review two approaches to compute the same vacuum-to-vacuum transition of $\chi$ in the presence of a $\phi$ background.
On the other hand, computation-wise, they look very different:
In the parametric-resonance approach, one needs to find the growing mode of equation of motion and integrate the growth factor $\lambda_k$ over the growing mode; on the other hand, in the Feynman-diagrammatic approach, one needs to identify all possible diagrams which can be cut into two pieces and evaluate the cut diagrams one by one.

To be more concrete, let us focus on narrow resonances.
As discussed above, it is known that growing modes appear only around $h=n_{\varphi}^{2}$ and each $n_{\varphi}$ contribution $\Gamma_{n_{\varphi}}$ can be interpreted as $n_{\varphi} \varphi \to 2 \chi$.
{Since there is no overlap in the phase space among different $n_{\varphi}$'s. the total decay rate is given by
$\Gamma = \sum_{n_{\varphi}} \Gamma_{n_{\varphi}}$.}
By introducing a perturbative parameter $\epsilon = h - n_{\varphi}^{2}$, one finds
\bal
\label{eq:momentum_chi}
k_{n_{\varphi}} = \frac{n_{\varphi} m_{\varphi}}{2} \beta_{n_{\varphi}} \sqrt{1 + \frac{\epsilon}{(n_{\varphi} \beta_{n_{\varphi}})^{2}}} \;, \quad \quad \beta_{n_{\varphi}} = \sqrt{1 - \left(\frac{2 m_{\chi}}{n_{\varphi} m_{\varphi}}\right)^{2}} \;.
\eal
Here $\beta_{n_{\varphi}}$ is the velocity of $\chi$ produced via $n_{\varphi} \varphi \to 2 \chi$ in the absence of background field.
Then, one finds
\bal
\Gamma_{n_{\varphi}} = \frac{n_{\varphi} m_{\varphi}^{4}\beta_{n_{\varphi}} }{64 \pi^{2}} \int d\epsilon \sqrt{1 + \frac{\epsilon}{(n_{\varphi} \beta_{n_{\varphi}})^{2}}} \lambda_{n_{\varphi}} (\epsilon; \theta) \;.
\eal
Here a growth factor $\lambda_{n_{\varphi}} (\epsilon; \theta)$ is an even function of $\theta$, since $\theta \to -\theta$ is absorbed by $z \to z + \pi/2$, which does not change the growing mode and factor. (Actually any complex phase of $\theta$ is absorbed in a similar way.)
{First using
\bal
\sqrt{1 + \frac{\epsilon}{(n_{\varphi} \beta_{n_{\varphi}})^{2}}} = \sum_{q=0}^{\infty} \frac{(-1)^{q-1}(2q-3)!!}{2^{q} q!} \frac{\epsilon^{q}}{(n_{\varphi} \beta_{n_{\varphi}})^{2q}} \;,
\eal
second performing the integral over $\epsilon$ (remember that the integration range of $\epsilon$ depends on $\theta$),
and third expanding the result in terms of $\theta$,
one sees the following double expansion:
\bal
\label{eq:double expansion}
\Gamma_{n_{\varphi}} = \sum_{p=1, q=0} \Gamma^{(p, q)}_{n_{\varphi}} \theta^{2p} \beta_{n_{\varphi}}^{1-2q} \,.
\eal
(Expansion in terms of $\beta_{n_{\varphi}}$ may not be quite physically meaningful, though.)}
One can obtain $\lambda_{n_{\varphi}} (\epsilon; \theta)$ only perturbatively as we see below, but for a given order of $\lambda_{n_{\varphi}} (\epsilon; \theta)$ one obtains all orders of $\beta_{n_{\varphi}}$ but with different powers of $p$ (remember $\epsilon^{q}$ in the above expansion).

Note that ``narrow-resonance regime'' in the parametric-resonance approach is equivalent to ``small amplitude'' (namely, perturbative) in the Feynman-diagrammatic approach.
Note that diagrams with more insertions give a larger contribution for large amplitude.
In the Feynman-diagrammatic approach, $p$ denotes the number of pairs of $\varphi_{+}$ (taking away energy of $m_\varphi$ from the $\chi$-loop) and $\varphi_{-}$ (bringing energy of $m_\varphi$) in external lines. (Diagrams with different numbers of $\varphi_{+}$ and $\varphi_{-}$ in external lines vanish because of energy conservation.)
The difference in numbers between $\varphi_{+}$ and $\varphi_{-}$ on one side of cut diagram is equal to $n_{\varphi}$ (the other side is equal to $- n_{\varphi}$).
This explains why only $p \geq n_{\varphi}$ gives finite $\Gamma_{n_{\varphi}}^{(p,q)}$.
The diagram with $n_{\varphi}$ $\varphi_{+}$'s next to each other in one side and $n_{\varphi}$ $\varphi_{-}$'s next to each other in the other side is the lowest-order contribution to $\Gamma_{n_{\varphi}}$ (see Fig.~2 of Ref.~\cite{Matsumoto:2007rd}).
There are in general more than one diagram contributing to $\Gamma_{n_{\varphi}}$ for given $p$.
{The summed amplitude depends on $\beta_{n_{\varphi}}$ implicitly through $m_{\chi}$ and thus can be expanded in terms of $\beta_{n_{\varphi}}$ to find $\Gamma_{n_{\varphi}}^{(p,q)}$.}
Note that unlike the parametric-resonance approach, where the $\beta_{n_{\varphi}}$ expansion also changes power of $\theta$ for a given order of $\lambda_{n_{\varphi}} (\epsilon; \theta)$, in the Feynman-diagrammatic approach, $\beta_{n_{\varphi}}$ expansion does not change a power of $\theta$.

One may wonder how a negative power of $\beta_{n_{\varphi}}$ appears.
To understand it, we start with noting that one (or both) side of cut diagram can diverge.
This occurs when more than one propagator is identical to each other and one of them is cut.
Since the cut one is taken to be on-shell, the others become on-shell, too, which leads to divergence.
This divergence actually cancels among different cuts of the diagram and finite contributions provide a negative power of $\beta_{n_{\varphi}}$ (see appendix~\ref{sec:apparent divergence}).
To circumvent such an apparent divergence, we apply a pinch technique. 
More specifically, when there are $m$ number of the same propagators, we rewrite it as
\bal
(k^{2} - m_{\chi}^{2} + i \epsilon)^{-m} = \lim_{\xi \to m_{\chi}^{2}} \frac{\partial^{m-1}}{\partial \xi^{m-1}} \frac{(-1)^{m-1}}{(m-1)!} (k^{2} - \xi + i \epsilon)^{-1} \;,
\eal
and then takes the limit and derivative after cutting the pinched diagrams.
The cut gives us the phase-space factor proportional to
\bal
\label{eq:B_kin}
{\rm B}_{\rm kin}(p^2,\xi_i,\xi_j)=
        \frac{1}{p^2}\sqrt{p^4-2(\xi_i+\xi_j)p^2+(\xi_i-\xi_j)^2} 
\eal
for $\sqrt{p^2}\gtrsim \xi_i^{1/2}+\xi_j^{1/2}$ (otherwise, 0).
It coincides with $\beta_{n_\varphi}$ when $\xi$'s coincide with $m_{\chi}^{2}$:
%${\rm B}_{\rm kin}(Q^2,m_\chi^2,m_\chi^2)=\sqrt{Q^4-2(\xi_i+\xi_j)Q^2+(\xi_i-\xi_j)^2}/Q^2$. 
%Hereafter, we have used the shorthand notation of the kinematical factor: 
\bal
\beta_{n_\varphi}= B_{\rm kin}(n_\varphi ^2m_\varphi^2,m_\chi^2,m_\chi^2)\;. 
\eal
Its derivative with respect to $\xi$ provides a negative power of $\beta_{n_{\varphi}}$.
Therefore, the number of derivatives acting on the cut propagators, determines the lowest power of $\beta$ appearing from a given diagram. (Note that not only the lower power term but also higher power terms can appear in general.)

\subsection{Double expansion ($\theta$ and $\beta$)} \label{sec:double expansion}
 \begin{figure}[t]
  \centering
  \includegraphics[scale=0.45]{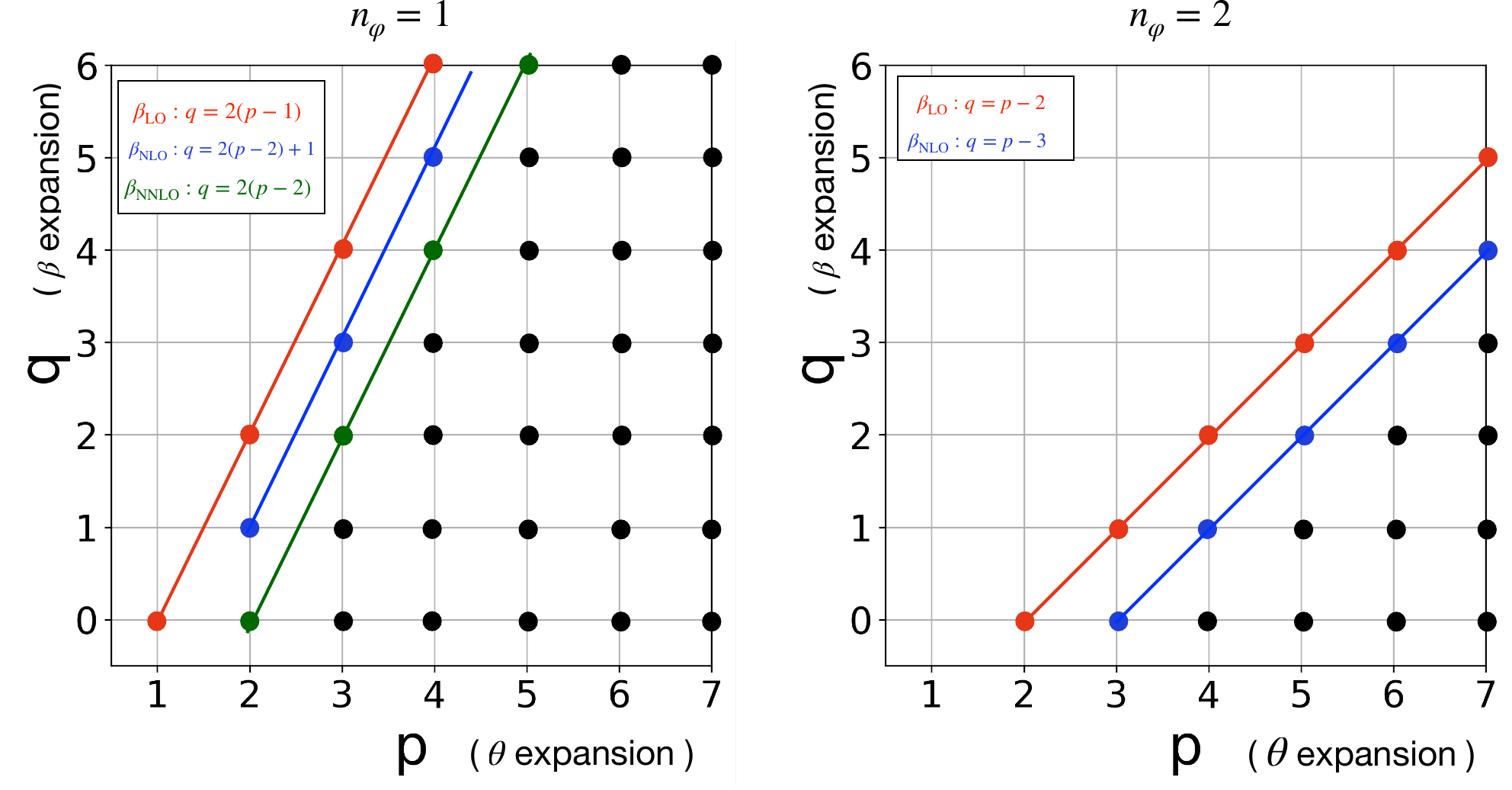}
   \caption{Schematic picture of $\beta$ and $\theta$ expansions. Left panel: the case of $n_\varphi=1$. Right panel: the case of $n_\varphi=2$. 
     }
  \label{FIG:lattice_points}
 \end{figure}

In the rest of this paper, we explicitly evaluate $\Gamma^{(p, q)}_{n_{\varphi}}$ both in the parametric-resonance approach (Sec.~\ref{sec:parametric}) and in the Feynman-diagrammatic approach (Sec.~\ref{sec:diagrammatic}) for $n_{\varphi} = 1 , 2$ and small $(p, q)$.
Fig.~\ref{FIG:lattice_points} depicts $(p, q)$ having non-zero contribution in black.
In the parametric-resonance approach, for a given order of $\lambda_{n_{\varphi}} (\epsilon; \theta)$ (LO, NLO, NNLO), one obtains all $(p, q)$ along the corresponding line.
We compute $p = 1, 2, 3$ for $n_{\varphi} = 1 , 2$ in the Feynman-diagrammatic approach, where all $q$ are obtained for each $p$ (vertically in the figure).
Only $p \geq n_{\varphi}$ gives finite $\Gamma_{n_{\varphi}}^{(p,q)}$ as discussed above in the Feynman-diagrammatic approach, this indicates that the leading order of $\beta$ expansion is of order of $\theta^{2 n_{\varphi}}$ in the parametric-resonance approach.
Actually, $\epsilon$ is of order of $\theta$ for $n_{\varphi} = 1$ and $\theta^{2}$ for $n_{\varphi} \geq 2$ (see Sec.~3.2 or Ref.~\cite{Yoshimura:1995gc}).
Also noting that at the leading order for $n_{\varphi} = 1$, $\lambda_{n_{\varphi}=1} (\epsilon; \theta)$ is an even function of $\epsilon$ and integration range is also symmetric around the origin, one understands $q = 2(p-1)$ is the largest $q$ for $n_{\varphi} = 1$ and given $p$; $q = p - n_{\varphi}$ is the largest $q$ for $n_{\varphi} \geq 2$ and given $p$.

\section{Results of the parametric-resonance approach}\label{sec:parametric}
\subsection{Methodology}
We here discuss the methodology of the calculation of the scalar condensate decay rate in the parametric-resonance approach. 
Since the decay rate is given by the growth factor $\lambda$, what one needs to do first is to derive the analytical expressions of the growth factor.    
To do this, we follow Ref.~\cite{Yoshimura:1995gc}. 
The starting point is the Mathieu equation Eq.~\eqref{eq:Mathieu}. 
The ansatz of the Mathieu equation Eq.~\eqref{eq:Matheiu_eq} is given by Eq.~\eqref{eq:special solution}. 
This periodic function can also be expressed by a Fourier series as
\bal
P(z)=\sum_{k=-\infty}^{\infty}c_k(z)e^{{i(n_\varphi+2k)}z} \;.
\eal
Inserting this in the ansatz $u(z)$ yields 
\bal \label{eq:ufre}
u=\sum_{k=-\infty}^{\infty}c_k(z)e^{\{\lambda+{i(n_\varphi+2k)}\}z} \;.
\eal
As above, we introduce the small perturbation for $h$, i.e., $h=2(E_{\chi}/m_\varphi)^2=n^2_\varphi+\epsilon$.
Substituting Eq.~\eqref{eq:ufre} into the Mathieu equation Eq.~\eqref{eq:Matheiu_eq}, 
one obtains the recurrence formula for $u_k$,
\bal \label{eq:thenkashiki}
\gamma_k c_k-\theta_+ c_{k+1}-\theta_- c_{k-1} =0\quad\quad\quad \mbox{with}\ \theta_{\pm}=\theta,
\eal
%with $\theta_{\pm}$ being $\theta_{\pm}=\theta$, 
where we have defined the factor $\gamma_k$ as
\bal
\gamma_k\equiv \{\lambda+i(n_\varphi+2k)\}^2+n^2_\varphi+\epsilon=\lambda^2+2i\lambda(n_\varphi+2k)-4k(n_\varphi+k)+\epsilon\;,
\eal

Eq.~\eqref{eq:thenkashiki} can also be expressed by tridiagonal matrix form as
\bal
\label{eq:tridiag}
\begin{pmatrix}
\gamma_N & -\theta_- &0 &\dots\\
-\theta_+& \gamma_{N-1} &-\theta_- & \dots\\
0&-\theta_+&\dots&\dots& \\
\dots&\dots&\dots&\dots&-\theta_-&\dots&\dots\\
\dots&\dots&\dots&-\theta_+&\gamma_0& \dots\\
\dots&\dots&\dots&\dots&\dots&\dots&-\theta_-&\dots&\dots&\\
\dots&\dots&\dots&\dots&\dots&-\theta_+&\gamma_{-n_\varphi}\\
\dots&\dots&\dots&\dots&\dots&\dots&\dots&\dots&\dots&\\
 \dots&\dots&\dots&\dots&\dots&\dots&\dots&\dots&\dots&-\theta_-\\
 \dots&\dots&\dots&\dots&\dots&\dots&\dots&\dots&-\theta_+&\gamma_{-N}
   \end{pmatrix}
   \begin{pmatrix}
    c_N\\
    c_{N-1}\\
    \dots\\
    \dots\\
    c_0\\
    \dots\\
    c_{-n_\varphi}\\
    \dots\\
    \dots\\
    _{-N}\\
   \end{pmatrix}
   =0 \;,
\eal
where one finally takes the limit of $N\to \infty$. 
One can divide this tridiagonal matrix into three blocks as
\bal
D&=
\begin{pmatrix}
  \gamma_N & -\theta_- &0 \\
  -\theta_+& \gamma_{N-1} &-\theta_- \\
  0&-\theta_+&\dots&\dots \\
  \dots&\dots&\dots&\dots&-\theta_-\\
  \dots&\dots&\dots&-\theta_+&\gamma_1\\
     \end{pmatrix}\;, \quad
E=
\begin{pmatrix}
  \gamma_{-1} & -\theta_- &0 \\
  -\theta_+& \gamma_{-2} &-\theta_- \\
  0&-\theta_+&\dots&\dots \\
  \dots&\dots&\dots&\dots&-\theta_-\\
  \dots&\dots&\dots&-\theta_+&\gamma_{-n_\varphi+1}\\
     \end{pmatrix}\;, \notag \\
F&=
\begin{pmatrix}
  \gamma_{-n_\varphi-1} & -\theta_- &0 \\
  -\theta_+& \gamma_{-n_\varphi-2} &-\theta_- \\
  0&-\theta_+&\dots&\dots \\
  \dots&\dots&\dots&\dots&-\theta_-\\
  \dots&\dots&\dots&-\theta_+&\gamma_{-N}\\
     \end{pmatrix}\;,
\eal
where $D$ is $N\times N$ matrix, $E$ is $(n_\varphi-1)\times (n_\varphi-1)$ matrix and $F$ is $(N-n_\varphi)\times (N-n_\varphi)$ matrix. 
Each block matrix can be solved separately. 
Solving the $D$ part of the tridiagonal matrix, one can obtain a recurrence relation for $c_1$ as
\bal \label{eq:c1}
c_1=\theta_-c_0 D^{-1}_{N,N} \;.
\eal
The same procedure for the $F$ part yields
\bal \label{eq:cnm1}
c_{-n_\varphi-1}=\theta_+ c_{-n_\varphi} F^{-1}_{1,1}\;.
\eal
From the $E$ part calculations, one obtains the recurrence relation for $c_{-1}$ and $c_{-n_\varphi+1}$ as
\bal \label{eq:cm1mn1}
c_{-1}&= E^{-1}_{1,1}\theta_+ c_0 + E^{-1}_{1,n_\varphi-1}\theta_- c_{-n_\varphi}\;, \\
c_{-n_\varphi+1}&= E^{-1}_{n_\varphi-1,1}\theta_+ c_0 +E^{-1}_{n_\varphi-1,n_\varphi-1}\theta_- c_{-n_\varphi}\;. 
\eal
Now we are ready to solve the recurrence relation for $c_0$ and $c_{-n_\varphi}$. 
From Eq.\eqref{eq:thenkashiki} and Eqs.~\eqref{eq:c1}-\eqref{eq:cm1mn1}, one can derive matrix equations for these coefficients as  
\bal 
\begin{pmatrix}
  \lambda^2+\epsilon +2in_\varphi\lambda -\theta_+ \theta_-(D^{-1}_{N,N}+E^{-1}_{1,1}) &-\theta_-^2 E_{1,n_\varphi-1}^{-1} \\
  -\theta_+^2 E_{n_\varphi-1,1}^{-1} & \lambda^2+\epsilon -2in_\varphi\lambda -\theta_+\theta_-(F^{-1}_{1,1}+E^{-1}_{n_\varphi-1,n_\varphi-1}) 
\end{pmatrix} 
\begin{pmatrix}
  c_0 \\
  c_{-n_\varphi}
\end{pmatrix} 
=0 \;.
\label{eq:matc0cmn}
 \eal
%where we have used $E_{1,1}^{-1}=E_{n_\varphi-1,n_\varphi-1}^{-1}$. 
We note that Eq.~\eqref{eq:matc0cmn} holds for $n_{\phi}\geq 2$. 
$n_\varphi=1$ is a special case, where the block matrix $E$ does not exist. 
In this case, one has the following matrix equation, 
\bal
\begin{pmatrix}
  \lambda^2+\epsilon +2i\lambda -\theta_+ \theta_-D^{-1}_{N,N} &-\theta_-  \\
 -\theta_+ & \lambda^2+\epsilon -2i\lambda -\theta_+\theta_-F^{-1}_{1,1} 
\end{pmatrix} 
\begin{pmatrix}
  c_0 \\
  c_{-1}
\end{pmatrix} 
=0 \;. \label{eq:mat_n=1}
\eal
Solving the determinant equal zero of the above coefficient matrix equations Eqs.~\eqref{eq:matc0cmn} and~\eqref{eq:mat_n=1}, one can calculate the relation between $\lambda$ and $\epsilon$. 
For the inverse matrix $D^{-1}$, $E^{-1}$ and $F^{-1}$, one can apply the formulae for the inverse matrix for the tridiagonal matrix (see, e.g., Ref.~\cite{tridiagonal}), which is discussed in the next section.

\subsection{Calculation of the matrix elements of the tridiagonal matrix}
The elements $F^{-1}_{1,1}$,$D^{-1}_{N,N}$, $E^{-1}_{1,1}$, $E^{-1}_{n_\varphi-1,n_\varphi-1}$, $E^{-1}_{1,n_\varphi-1}$ and $E^{-1}_{n_\varphi-1,1}$ appear in Eq.~\eqref{eq:matc0cmn}. 
Here we discuss how we calculate these matrix elements by using the properties of the tridiagonal matrix. 

%To describe this, we first fix our convention. 
We discuss the derivations of the concrete expressions of these matrix elements, we first give the formula of the inverse of tridiagonal matrix. 
One considers the general form of a $n\times n$ tridiagonal matrix as
\bal
T=
\begin{pmatrix}
  a_1 & b_1&0 \\
  c_1& a_2 &b_2 \\
  0&c_2&\dots&\dots \\
  \dots&\dots&\dots&\dots&b_{n-1}\\
  \dots&\dots&\dots&c_{n-1}&a_n\\
     \end{pmatrix}\;.
\eal
The inverse of $T$ is then given by~\cite{tridiagonal}
\bal
T^{-1}_{i,j}
=
\begin{cases}
    (-1)^{i+j}b_i\dots b_{j-1} \tilde{\theta}_{i-1}\frac{\tilde{\phi}_{j+1}}{\tilde{\theta}_n}& (i<j) \\
    \tilde{\theta}_{i-1}\frac{\tilde{\phi}_{j+1}}{\tilde{\theta}_n} & (i=j) \\
    (-1)^{i+j}c_j\dots c_{i-1} \tilde{\theta}_{j-1}\frac{\tilde{\phi}_{i+1}}{\tilde{\theta}_n}& (i>j)\;,
  \end{cases}
\label{eq:inverse_tridiagonal}
\eal
%From the original tridiagonal matrix in the Eq.~\eqref{eq:tridiag}, one defines the following quantities
The quantities $\tilde{\phi}$ and $\tilde{\theta}$ satisfy the following recurrence relations:
\bal
\tilde{\theta}_i=a_i\tilde{\theta}_{i-1}-b_{i-1} c_{i-1} \tilde{\theta}_{i-2}\;, (i=2,3,\dots, n)\;, \notag \\ \label{eq:phi_rec}
\tilde{\phi}_i= a_i \tilde{\phi}_{i+1}-  b_{i} c_{i}\tilde{\phi}_{i+2}\;, (i=n-1,\dots, 1)\;,
\eal
where initial conditions for these quantities are given by $\tilde{\theta}_0=1$, $\tilde{\theta}_1=a_1$, $\tilde{\phi}_{n+1}=1$ and $\tilde{\phi}_{n}=a_n$. 

We first derive general expressions for the elements of the inverse of the tridiagonal matrices $F,\ D$, and $E$ that are valid for any $\theta$ order. 
Let us focus on $F^{-1}_{1,1}$, introducing the ratio $x_i\equiv\tilde{\phi}_{i+1}/\tilde{\phi}_i$, for which we note that the recurrence relation for $x_i$ is obtained from Eq.~\eqref{eq:phi_rec},  
$1=a_ix_i- b_ic_i x_ix_{i+1}$.
%\bal
% 1=\gamma_ix_i-\theta_+ \theta_- x_i x_{i+1} .
%\eal
Using the formulae of the inverse for the tridiagonal matrix Eq.~\eqref{eq:inverse_tridiagonal}, $F^{-1}_{1,1}$ is given by
\bal
F_{1,1}^{-1}=\tilde{\theta}_0(F)\frac{\tilde{\phi}_2(F)}{\tilde{\theta}_{N-n_\varphi}(F)}=\frac{\tilde{\phi}_2(F)}{\tilde{\theta}_{N-n_\varphi}(F)}=\frac{\tilde{\phi}_2(F)}{\tilde{\phi}_1(F)}= x_1(F),
\eal
where in the second equality, we have used the initial condition of $\tilde{\theta}_0$  and in the third equality, we have used $\tilde{\theta}_n$=$\tilde{\phi}_1$~\cite{tridiagonal}.
The inverse of the matrix $D$ can be evaluated by introducing another ratio parameter, $y_i\equiv{\tilde{\theta}_{i-1}}/{\tilde{\theta}_i}$. 
$D_{N,N}^{-1}$ is given by  
\bal
D_{N,N}^{-1}=y_N(D) \;.
\eal
Similarly, for the matrix $E$, one obtains
\bal
E_{1,1}^{-1}=x_1(E)\;,\quad E_{n_\varphi-1,n_\varphi-1}^{-1}=y_{n_\varphi-1}(E)\;, \notag \\
E_{1,n_\varphi-1}^{-1}=(-1)^{n_\varphi}\theta_+^{n_\varphi-2}\frac{1}{\tilde{\theta}_{n_\varphi-1}(E)} \;,\quad
E_{n_\varphi-1,1}^{-1}&=(-1)^{n_\varphi}\theta_-^{n_\varphi-2}\frac{1}{\tilde{\theta}_{n_\varphi-1}(E)} \;.
\eal
$E_{1,n_\varphi-1}^{-1} = E_{n_\varphi-1,1}^{-1}$ for $\theta_+ = \theta_-$.
We denote quantities of which tridiagonal matrix by an argument.

In the following, we expand in $\theta$ and explicitly compute more concrete expressions for the inverse of tridiagonal matrices. 
Since we will apply the above formula of $T^{-1}$ to the block matrices $D$, $E$, and $F$, the parameter $b_i$ and $c_i$ are commonly identified by $\theta$. 
Hence, by expanding $\theta$ in the $x$, i.e., $x_i=x_i^{\rm LO}+x_i^{\rm NLO}+\dots$, $x_1$ is written by
\bal
x_1^{\rm LO}=\frac{1}{a_1} \;,\quad
% x_1^{(1)}=\frac{\theta^2}{a_1^2 a_2} \;\;.
 x_1^{\rm NLO}=\frac{\theta_+\theta_-}{a_1^2 a_2} \;\;.
\eal
With $a_1(F)=\gamma_{-n_\varphi-1},\; a_2(F)=\gamma_{-n_\varphi-2}$ for the matrix $F$, one obtains 
\bal
F_{1,1}^{-1}\simeq\frac{1}{\gamma_{-n_\varphi-1}}
\left( 1+\frac{\theta_+\theta_-}{\gamma_{-n_\varphi-1} \gamma_{-n_\varphi-2}} \right).
\eal
The ratio $y_i$ is also expanded by $\theta$ as $x_i$. 
The matrix element $D_{N,N}^{-1}$ is given by  $D_{N,N}^{-1}=y_N(D)$. 
The recurrence relation for $y_N$ can be obtained from Eq.~\eqref{eq:phi_rec} and used for the derivation of $y_N^{\rm LO}$ and $y_N^{\rm NLO}$.
Thus, including higher order terms in the $\theta$ expansion, $D_{N,N}^{-1}$ is given by
\bal
 D_{N,N}^{-1}\simeq\frac{1}{\gamma_1}\left(1+\frac{\theta_+\theta_-}{\gamma_1  \gamma_2}\right) \;.
\eal
Similarly, for the matrix $E$, the inverse matrix can be calculated by 
\bal
E_{1,1}^{-1}&\simeq\frac{1}{\gamma_{-1}}
 \left( 1+\frac{\theta_+ \theta_-}{\gamma_{-1} \gamma_{-2}} \right)\;, \\
E_{n_\varphi-1,n_\varphi-1}^{-1}&\simeq\frac{1}{\gamma_{-n_\varphi+1}}\left(1+\frac{\theta_+ \theta_-}{\gamma_{-n_\varphi+1}  \gamma_{-n_\varphi+2}}\right)\;, \\ 
E_{1,n_\varphi-1}^{-1}&\simeq(-1)^{n_\varphi}\theta_+^{n_\varphi-2}\left(\prod_{k=1}^{n_\varphi-1}\frac{1}{\gamma_{-k}}\right), \\
E_{n_\varphi-1,1}^{-1}&\simeq(-1)^{n_\varphi}\theta_-^{n_\varphi-2}\left(\prod_{k=1}^{n_\varphi-1}\frac{1}{\gamma_{-k}}\right)\;.
\eal
where $\tilde{\theta}_{n_{\varphi} - 1}^{\rm LO} (E) = \prod_{k=1}^{n_\varphi-1} \gamma_{-k}$ is used.

Neglecting higher order terms in $\epsilon$ and $\lambda$, $\gamma_k$ can be reduced as
\bal
\gamma_k&\simeq-4k(n_\varphi+k) \;,\label{eq:gammak}\\
\gamma_0&\simeq\epsilon+2in_\varphi \lambda, ~~~(k=0)\;, \label{eq:gamma0} \\
\gamma_{-n}&\simeq\epsilon-2in_\varphi \lambda, ~~~(k=-n_\varphi). 
\label{eq:gammamn}
\eal
We note that one holds $\gamma_{k}=\gamma_{-k-n_\varphi}$ and thus $F_{1,1}^{-1} = D_{N,N}^{-1}$ $E_{1,1}^{-1}=E_{n_\varphi-1,n_\varphi-1}^{-1}$ and $E_{1,n_\varphi-1}^{-1} = E_{n_\varphi-1,1}^{-1}$
at the leading order. 
Then, the determinant equal to zero gives
\bal
\lambda^{(0)} &\simeq \frac{1}{2n_{\varphi}} \sqrt{\theta^{4} E_{1,n_\varphi-1}^{-1} E_{n_\varphi-1, 1}^{-1} - \left( \epsilon - \theta^{2} (D_{N,N}^{-1} + E_{1,1}^{-1}) \right)^{2} } \notag \\
&= \frac{1}{2n_{\varphi}} \sqrt{\left( \frac{\theta^{n_{\varphi}}}{4^{n_{\varphi}-1} [(n_{\varphi}-1)!]^{2}} \right)^{2} - \left( \epsilon - \frac{\theta^{2}}{2 (n_{\varphi}^{2} -1)} \right)^{2}}
\eal
for $n_{\varphi} \geq 2$ and
\bal
\lambda^{(0)} &\simeq \frac{1}{2} \sqrt{\theta^{2} - \left( \epsilon - \theta^{2} D_{N,N}^{-1} \right)^{2}} = \frac{1}{2} \sqrt{\theta^{2} - \left( \epsilon + \frac{\theta^{2}}{8} \right)^{2}}
\eal
for $n_{\varphi} =1$ ($\theta^2/8$ in the right-most side can be dropped at the leading order).

\subsection{Analytical results with higher order terms}
As discussed above, the growth factor $\lambda$ can be expressed in terms of $\theta$ and $\epsilon$ by the determinant of Eq.~\eqref{eq:matc0cmn} and \eqref{eq:mat_n=1} for $n_\varphi\geq2$ and $n_\varphi=1$, respectively. 

In the case of $n_\varphi=1$, the LO expression of $\lambda$ is simply given by 
$\lambda^{(0)}=\sqrt{\theta^2-\epsilon^2}/2$. 
This yields the decay rate of $\varphi$ at the LO as~\cite{Matsumoto:2007rd}
\bal
\Gamma_{n_\varphi=1}^{\rm LO}&=\frac{m_\varphi^4}{128\pi^2}\int^\theta_{-\theta}d \epsilon\sqrt{\beta_1^2+\epsilon}\sqrt{\theta^2-\epsilon^2} \notag \\
&=-\frac{m_\varphi^4}{16\pi}\sum_{p=1}^\infty\beta^{5-4p}_1\left(\frac{\mu A_\varphi}{2m_\varphi^2}\right)^{2p}
\frac{(4p-7)!!}{(p-1)!p!} \;,
\label{eq:nphi=1 LO result}
\eal
where
\bal
\int_{0}^{\frac{\pi}{2}} dx \cos^{2n}x \sin^2x = \frac{\pi}{4} \frac{(2n-1)!!}{2^n (n+1)!}
\eal
for integer $n$ would be useful.
In the following subsections, we discuss the derivation of the decay rates of $\varphi$ including higher-order terms in the $\theta$ expansions, focusing on $n_\varphi=1$ and $n_\varphi=2$. 

\subsubsection{$n_\varphi=1$}
We turn into the derivation of analytical expressions of the growth factor with NNLO terms of $\theta$ expansions, by which one can derive terms up to NNLO in $\beta$ expansion for a given order of $\theta$. 
From the determinant of Eq.~\eqref{eq:mat_n=1}, one obtains the quadratic equation of $\lambda$,
\bal
0&\simeq\Big[  (\lambda^{(1)})^2 +\epsilon -\theta^2\Big\{-\frac{1}{8}-\frac{\epsilon}{64}\Big\}\Big]^2+4\lambda^2\left(1+\frac{3}{64}\theta^2\right)^2-\theta^2
\eal
with the up-to-NLO expression of $\lambda$ being
\bal
(\lambda^{(1)})^2=\frac{1}{4}\left[\theta^2-\left\{\frac{1}{4}\theta^2-\epsilon^2+\epsilon+\frac{1}{8}\theta^2 \right\}^2\right]\;,
\eal
which is obtained from the determinant with $(\lambda^{(0)})^{2}$.
We have used the concrete expression of the tridiagonal matrix elements:
\bal
D_{N,N}^{-1}&\simeq-\frac{1}{8}-\frac{\epsilon}{64}-\frac{3i\lambda}{32}\;, \notag \\
F_{1,1}^{-1}&\simeq-\frac{1}{8}-\frac{\epsilon}{64}+\frac{3i\lambda}{32}\;.
\eal
This yields the up-to-NNLO expression of $\lambda$ as
\bal
\lambda^{(2)}
\simeq\frac{1}{16}\left(1-\frac{3}{32}\theta^2\right)^{1/2}\prod_{i=1}^3 \sqrt{\epsilon-\alpha_i}\sqrt{\epsilon-\beta_i},
\eal
where $\alpha_i$ and $\beta_i$ $(i=1,2,3)$ are the solutions of $f_{+}=0$ and $f_{-}=0$, respectively, where
\bal
f_\pm=\theta \pm\frac{1}{8} (\epsilon -2) \epsilon ^2\pm\frac{1}{64}
   \theta ^2 (24-11 \epsilon )\pm\epsilon\;.
\eal
If one expands $\alpha_i$ and $\beta_i$ by $\theta$ (not in the actual computation), they are given by
$\alpha_3\simeq-\theta-\theta^2/8+\theta^3/64$, $\alpha_2\simeq z_1+{\theta}(z_2/2)+\theta^2(z_3/16)$, 
% $\alpha_1\simeq z_1^\ast+\frac{\theta}{14}z_2^\ast$
$\beta_3\simeq \theta-\theta^2/8-\theta^3/64$, 
$\beta_2\simeq z_1^\ast-{\theta}(z_2^\ast/2)+\theta^2(z_3^\ast/16)$, 
$\alpha_1\simeq \alpha_2^\ast$, and $\beta_1\simeq \beta_2^\ast$, 
% $\beta_1\simeq z_1^\ast-\frac{\theta}{14}z_2^\ast$ 
with $z_1$, $z_2$, and $z_3$ being $z_1=1+i\sqrt{7}$, $z_2=1+i/\sqrt{7}$, and $z_3=1-30i/(7\sqrt{7})$.  

The parametric resonance occurs in 
%$\lambda \in \mathbb{R}_{>0}$, so that 
the range of $\alpha_3<\epsilon<\beta_3$. 
The decay rate of the scalar condensate is given by 
\bal
\Gamma_{n_\varphi=1}
&\simeq\frac{m_\varphi^4}{128\pi^2} 
  \left(1-\frac{3}{32}\theta^2\right)^{1/2} \notag \\
&\times\int^{\beta_3}_{\alpha_3}d\epsilon
\sqrt{\beta_1^2+\epsilon}
  \sqrt{\left(1+\frac{\tilde{f}_+}{8}\right) \left(1+\frac{\tilde{f}_-}{8}\right)}\sqrt{(\epsilon-\alpha_3) (\beta_3-\epsilon)}
\eal
with $\tilde{f}_{\pm}$ being $\tilde{f}_{+}=(\epsilon-\alpha_1)(\epsilon-\alpha_2)-8$ and $\tilde{f}_{-}=(\epsilon-\beta_1)(\epsilon-\beta_2)-8$. 

One can expand the integrand by, e.g., $\alpha_3$ and $\beta_3$ which are ${\cal O}(\theta)$. 
One then performs the integration with respect to $\epsilon$ by replacing the integration variable, $\epsilon=\epsilon^\prime+(\alpha_3+\beta_3)/2$. 
After these procedures one obtains the terms with the form Eq.~\eqref{eq:double expansion}.
%$\Gamma_{n_\varphi}\ \ni \Gamma^{(q,p)}_{n_\varphi}\beta^q \theta^{2p}$. 
As discussed in Sec.~\ref{sec:comparison}  (e.g., see Fig.~\ref{FIG:lattice_points}), in the parametric-resonance approach, one obtains lower-order terms in the $\beta$ expansion for given $p$. 
We specify $q=2p-2$ and then obtain the lowest-order terms of $\beta$ expansion, which is given in Eq.~\eqref{eq:nphi=1 LO result}. 
Similarly, we obtain the NLO and the NNLO expressions of the decay rates;
\begin{equation}
\boxed{
\begin{aligned}
\Gamma_{n_\varphi=1}^{\rm NLO}&=-\frac{m_\varphi^4}{16\pi}\sum_{p=2}^\infty\beta^{7-4p}
\left(\frac{\mu A_\varphi}{2m_\varphi^2}\right)^{2p}
\frac{(1+p)(4p-9)!!}{p!(p-2)!} \;, \\ 
\Gamma_{n_\varphi=1}^{\rm NNLO}&=-\frac{m_\varphi^4}{32\pi}\sum_{p=2}^\infty\beta^{9-4p}
\left(\frac{\mu A_\varphi}{2m_\varphi^2}\right)^{2p}
\frac{\{p(p-1)^2-6\}(4p-11)!!}{p!(p-2)!}.    
\end{aligned}
}
\end{equation}
Here we extract terms satisfying $q=2p-3$ and $q=2p-4$ for $\Gamma_{n_\varphi=1}^{\rm NLO}$ and  $\Gamma_{n_\varphi=1}^{\rm NNLO}$, respectively.

\subsubsection{$n_\varphi=2$}
Similar to the case of $n_\varphi=1$, one can calculate the growth factor including higher-order terms of $\theta$ and derive the analytical results of the decay rate. 
For the case of $n_\varphi=2$, we derive the decay rate up to the NLO terms in the $\beta$ expansion for a given order of $\theta$. 

Deriving the concrete expression of tridiagonal matrix elements up to the order ${\cal O}({\theta^2})$, from Eq.~\eqref{eq:matc0cmn}, one can derive the explicit up-to-NLO expression formula of $\lambda$ as
\bal
\lambda^{(1)}&\simeq\frac{1}{4}\left(1-\frac{\theta^2}{36}\right)^{1/2}
\Bigg[\frac{\theta^2}{4}\left(1-\frac{\epsilon}{4}\right) 
+\Big\{\epsilon-\frac{\epsilon^2}{16}-\frac{\theta^2}{6}-\frac{13}{144}\theta^2\epsilon+\frac{11}{4608}\theta^4 \Big\}\Bigg]^{1/2} \notag \\
&\times
\Bigg[\frac{\theta^2}{4}\left(1-\frac{\epsilon}{4}\right) 
-\Big\{\epsilon-\frac{\epsilon^2}{16}-\frac{\theta^2}{6}-\frac{13}{144}\theta^2\epsilon+\frac{11}{4608}\theta^4 \Big\}\Bigg]^{1/2} \;,
\eal
with LO lambda expression being $4\lambda^{(0)}=\{\theta^4/16-(\epsilon-\theta^2/6)^2\}^{1/2}$.
Each of the square brackets contains a quadratic expression in $\epsilon$. 
By using the solutions of these quadratic equations, i.e., $\alpha_{i}$ and $\beta_{i}$ $(i=1,2)$, 
$\lambda^{(1)}$ can be further reduced as follows;
\bal
\label{eq:lambda_nphi=2}
\lambda^{(1)}=
\frac{1}{4}\left(1-\frac{\theta^2}{36}\right)^{1/2}\sqrt{(\epsilon-\alpha_2) (\beta_2-\epsilon)}\sqrt{(1+\alpha_1-16-\epsilon)(1+\beta_1-16-\epsilon)}. 
\eal
where the $\theta$ expansion of $\alpha_{1,2}$ and $\beta_{1,2}$ are given by
$\alpha_2\simeq-\theta^2/12$, $\beta_2\simeq-5\theta^2/12$, $\alpha_1\simeq16+19\theta^2/36$ and  $\beta_1\simeq16+73\theta^2/36$. 

The range of $\epsilon$ is determined from Eq.~\eqref{eq:lambda_nphi=2} as $\alpha_2<\epsilon<\beta_2$. 
The decay rate for $n_\varphi=2$ with the NLO growth factor is then given by
\bal
\Gamma_{n_\varphi=2}&\simeq\frac{m_\varphi^4}{128\pi^2}\int_{\alpha_2}^{\beta_2}d \epsilon\sqrt{1-\frac{\theta^2}{36}}\sqrt{\beta_2^2+\frac{\epsilon}{4}} 
\sqrt{(\epsilon-\alpha_2) (\beta_2-\epsilon)} \notag \\
&\times\sqrt{(1+\alpha_1-16-\epsilon)(1+\beta_1-16-\epsilon)}
\eal
By performing the integration with respect to $\epsilon$ and extracting the terms that satisfy specific relations between $q$ and $p$, the LO and NLO terms in the $\beta$ expansion can be obtained as
\begin{equation}
\label{eq:Gam_nphi=2_para}
\boxed{
\begin{aligned}
\Gamma_{n_{\varphi}=2}^{\rm LO}&=\sum_{p=2}\frac{2^{-4(1+p)}3^{2-p}m_\varphi^4\beta_2^{5-2p}\theta^{2p} }{\pi(p-2)!(5-2p)!!}\ _2F_1\left(\frac{3-p}{2},1-\frac{p}{2};2;\frac{9}{4}\right), \\  
\Gamma_{n_{\varphi}=2}^{\rm NLO}&=\sum_{p=3}\frac{(-1)^{1+p}(2p-9)!!m_\varphi^4\beta_{2}^{7-2p}\theta^{2p} }{3^{1+p}\pi p!}
\Bigg[4\{315+p(789+2p(-509+144p)) \}\ _2F_1\left(\frac{3-p}{2},2-\frac{p}{2};1;\frac{9}{4}\right) \\
&-5(-3+p)(51+37p)\ _2F_1\left(\frac{5-p}{2},2-\frac{p}{2};1;\frac{9}{4}\right) \Bigg],
\end{aligned}
}
\end{equation}
where $\ _2F_1(a,b;c;d)$ is the hypergeometric function. 
To derive Eq.~\eqref{eq:Gam_nphi=2_para}, we have set $q=p-2$ and $q=p-3$, respectively.

Before closing this section, let us briefly discuss the hypergeometric function to understand the above results better. 
The definition is given by 
\bal
\ _2F_1(a,b;c;d)=\frac{\Gamma(c)}{\Gamma(a) \Gamma(b)}\sum_{s=0}^\infty\frac{\Gamma(a+s) \Gamma(b+s)}{\Gamma(c+s)s!}z^s\;.
\eal
From the definition, obviously, one holds $_2F_1(b,a;c;d)=$$_2F_1(a,b;c;d)$. 
Another important property of this function is that in the case of $a$ (or $b$) is a non-positive integer, $_2F_1(a,b;c;d)$ is given by a finite number of terms, i.e.,
\bal
_2F_1(-m,b;c;z)=\frac{\Gamma(c)}{\Gamma(b)}\sum_{s=0}^m(-1)^s \begin{pmatrix} m \\ s\end{pmatrix}
\frac{\Gamma(b+s)}{\Gamma(c+s) s!}z^s\;, 
\eal
with $m=0,1,2,\dots$

\section{Results of the Feynman-diagrammatic approach}\label{sec:diagrammatic}
\subsection{Evaluation of bubble diagrams}

\begin{figure}[tbhp]
  \centering 
  \includegraphics[scale=0.45]{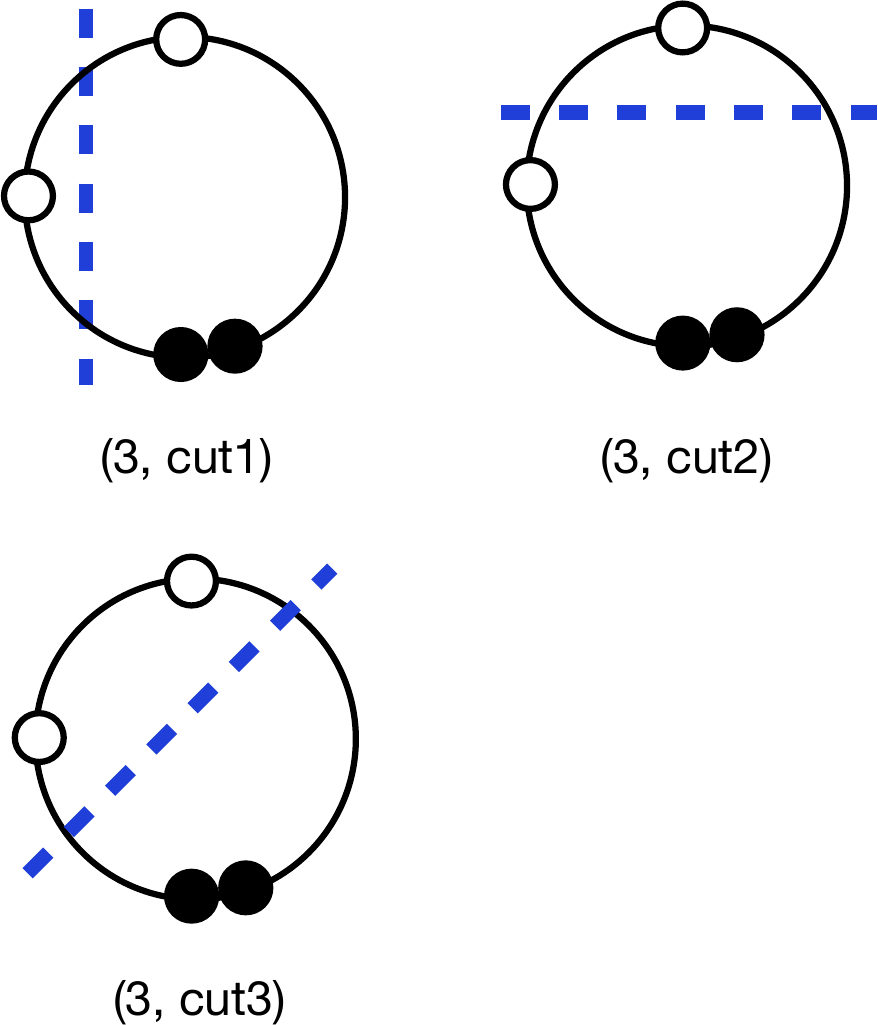}
   \caption{Possible cuts for the ``pinched'' diagram for the one of $p=2$ graphs.
   Diagrams 2-1 and 2-2 contribute to the $n_\varphi=1$ process, while Diagram 2-3 contributes to the $n_\varphi=2$ process. 
     }
  \label{FIG:p2_pinched}
 \end{figure}

In the Feynman-diagrammatic approach, the scalar condensate decay rate is obtained by evaluating bubble diagrams. 
We here describe the procedure to evaluate the bubble diagrams. 
It is constructed in three steps. 

In the first step, one ``pinches'' bubble diagrams.
Suppose that one has a bubble diagram with amplitude $i{\cal M}$ with $2p$ insertion of the black ($\varphi_+$ and thus outgoing $Q=(m_\varphi,{\bf 0})$) and white ($\varphi_-$ and thus incoming $Q$) circles. One then has $2p$ propagators in the diagram, for which the momentum is labeled by $k_i$ $(i=1,\dots,2p)$. 
Noting the appearance of the same propagators, propagators in the bubble diagram $i{\cal M}$ can be reduced as
\begin{align}
(k_1^2-m_\chi^2+i\epsilon)^{-m_1}&(k_2^2-m_\chi^2+i\epsilon)^{-m_2}\dots (k_n^2-m_\chi^2+i\epsilon)^{-m_n} \notag\\
&=\lim_{\xi_1\to m_\chi^2} \lim_{\xi_2\to m_\chi^2}\dots \lim_{\xi_n\to m_\chi^2}\frac{1}{(m_1-1)!(m_2-1)!\ \dots(m_n-1)!}  \notag \\
&\times\frac{\partial^{m_1-1}}{\partial \xi_1^{{m_1-1}}}
\frac{\partial^{m_2-1}}{\partial \xi_2^{{m_2-1}}} \dots
\frac{\partial^{m_n-1}}{\partial \xi_n^{{m_n-1}}} 
\prod_{j=1}^{n}(k_j^2-\xi_j+i\epsilon)^{-1}\;,
%(k_1^2-\xi_1+i\epsilon)^{-1}(k_2^2-\xi_2+i\epsilon)^{-1}\dots(k_n^2-\xi_n+i\epsilon)^{-1} \;,
\end{align}
where the indices $m_i$ ($i=1,2\dots,n$ running over only different loop momenta) satisfy $\sum_{i=1}^{n}m_i=2p$ and we define $\frac{\partial^0}{\partial \xi_i^0}f(\{\xi_i\})=f(\{\xi_i\})$.

In the second step, one applies cutting rules. The decay rate of the scalar condensate is derived from the imaginary part of the bubble diagrams, which can be evaluated by the cutting rules.
Thus, we replace a pair of the propagators by the delta function as 
\begin{align}
  \prod_{j=1}^{n}(k_j^2-\xi_j+i\epsilon)^{-1}\to
  \prod_{\substack{j=1\\ j\neq\ell\neq m}}^{n}(k_j^2-\xi_j+i\epsilon)^{-1}(-2i\pi)^2\delta(k^2_\ell-\xi_\ell) \delta(k^2_m -\xi_m)\;,
\end{align}
and then sum up all possible cuts.
As a concrete example, the pinched diagrams of $p=2$ are shown in Fig.~\ref{FIG:p2_pinched}. 
For Diagram 2, one cut diagram contributing $n_\varphi=1$ exists. Since Diagram 3 possesses three propagators, three cut diagrams exist.
The first and second diagrams contribute to $n_{\varphi}=1$ and the third diagram contributes to $n_{\varphi}=2$. 
Then, one can use
\bal 
\label{eq:int formula}
\int \frac{d^4 k}{(2\pi)^4} \delta (k_i^2-\xi_i) \delta (k_j^2-\xi_j) &=
\int \frac{d^4 k}{(2\pi)^4} \frac{d^4 k'}{(2\pi)^4} (2\pi)^4 \delta^{4} (k_i-k_j - k - k') \delta (k^2-\xi_i) \delta (k'^2-\xi_j) \notag \\
&= \frac{1}{32\pi^3}B_{\rm kin}((k_i-k_j)^2,\xi_i,\xi_j)\;,
\eal
%with $k_i$ and $k_j$ being the loop momenta of $\chi$. 
where the kinematical function ${\rm B}_{\rm kin}$ is given by Eq.~\eqref{eq:B_kin}. 
(Precisely speaking, the first equality holds only for energy of $k_i-k_j$ is positive; if negative, replace it by $k_j-k_i$).
The first equality is reduced to the phase-space integration.
%, leading to the decay rate of the scalar condensate. 
{
In this way, the differential decay rate $d\Gamma$ can also be obtained, which we will discuss in Sec.~\ref{sec:conclusion}, though we perform the phase-space integration here to obtain the decay rate.
}
After these procedures, the imaginary part is obtained from the 
discontinuity, ${\rm Disc}.(i{\cal {M}|_{\rm pinched}})=-2{\rm Im}{\cal M|_{\rm pinched}}$. 
%For a certain pinched bubble diagram, one calculates the sum of possible cuts. 

In the last step, one performs the derivatives and takes the limits with respect to $\xi_i$. Then, the imaginary part of the bubble diagrams can be computed. In the generic case, the imaginary part of the bubble diagram is given by
\begin{align}
  i{\rm Im}{\cal M}=  
\lim_{\xi_1\to m_\chi^2} \lim_{\xi_2\to m_\chi^2} \dots \lim_{\xi_n\to m_\chi^2} \frac{\partial^{m_1-1}}{\partial \xi_1^{{m_1-1}}}
\frac{\partial^{m_2-1}}{\partial \xi_2^{{m_2-1}}} \dots
\frac{\partial^{m_n-1}}{\partial \xi_n^{{m_n-1}}} 
i{\rm Im}{\cal M|_{\rm pinched}} \;. 
\end{align}
In the end, one can obtain the decay rate of the scalar condensate by plugging the result of ${\rm Im}{\cal M}$ into 
\begin{align}
    \Gamma=2{\rm Im}{\cal M} .
\end{align}
Note that the delta function for the energy-momentum conservation, which is multiplied by the amplitude, turns into $L^{3} T$ since a bubble diagram trivially satisfies the energy-momentum conservation.

\subsection{$p=1$}
\begin{figure}[t]
  \centering
  \includegraphics[scale=0.6]{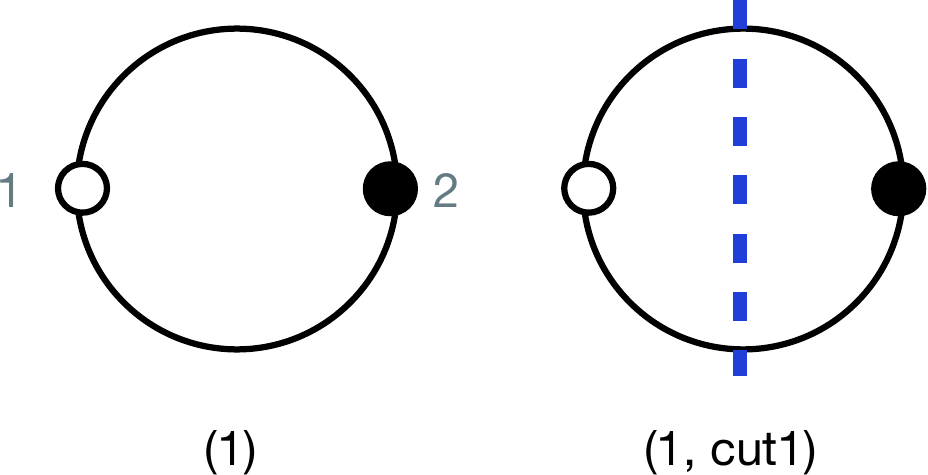}
   \caption{$p=1$ diagram and possible cut contributing to $n_\varphi=1$.
     }
  \label{FIG:p=1}
 \end{figure}
We first calculate the leading order contributions in the small $\theta$ expansions for $n_\varphi=1$, namely, the $p=1$ diagram shown in Fig.~\ref{FIG:p=1}. As can be understood from the momentum flow of the diagram, no common momentum exists. 
Applying the Feynman rules, the amplitude ${\cal M}_1$ is given by
\bal
{{\cal M}}_1&=
S_{F,1}\left(\frac{\mu A_\varphi}{2}\right)^4
\int \frac{d^4 k}{(2\pi)^4} (k^2-m_{\chi}^2+i\epsilon)^{-1} \{(k+Q)^2-m_{\chi}^2+i\epsilon\}^{-1}\;,
\label{eq:dia1}
\eal
where $S_{F,1}$ denotes the symmetric factor for Diagram ${\cal M}_1$, which is $S_{F,1}=1/2$ in this case. 
The symmetric factor can be calculated by counting the number of symmetries possessed by the diagram under consideration.
In the case of Diagram 1, there is one rotational symmetry (just unit element) and two reflection symmetries (unit element and reflection around the horizontal line connecting circles 1\&2), so that the symmetric factor becomes 1/2.
One can calculate the imaginary part of the ${{\cal M}}_1$ with the cutting rule as
\bal
{\rm Im}{\cal M}_{1}=
2\pi^2 \left(\frac{\mu A_\varphi}{2}\right)^2
\frac{1}{32\pi^3}B_{\rm kin}(Q^2,m_\chi^2,m_\chi^2)\;.
\eal
The decay rate is given by
\bal
\label{eq:nphi=1 p=1 result}
\boxed{
\Gamma_{n_\varphi=1}^{p=1}=\frac{m_\varphi^4}{16\pi}\left(\frac{\mu A_\varphi}{2m_\varphi^2}\right)^2\beta_{1}
} \;.
\eal
\subsection{$p=2$}
\begin{figure}[t]
  \centering
  \includegraphics[scale=0.6]{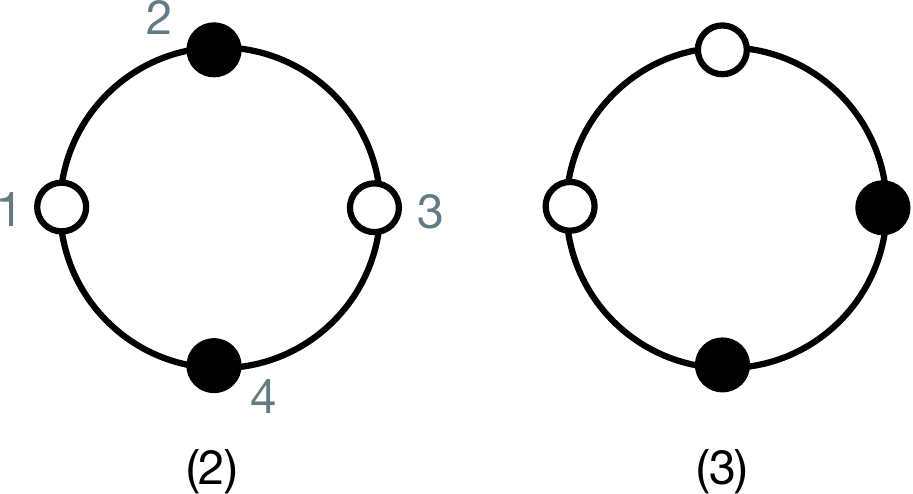}
   \caption{$p=2$ diagrams.
     }
  \label{FIG:p=2}
 \end{figure}

We here derive the analytical results of the NLO $\theta$ contribution from the bubble diagrams with $p=2$. 
As mentioned above, there are three possible cuts for $p=2$ Diagram 3 in Fig.~\ref{FIG:p2_pinched}, which is denoted by ${\cal M}_{3\mbox{-}i}$ $(i=1,2,3)$. 
Applying the pinch technique, the amplitude for Diagram 3 shown in Fig.~\ref{FIG:p=2} is given by
\bal
i {\cal M}_3&
=\lim_{\xi\to m_{\chi}^2}\frac{\partial }{\partial \xi}\widetilde{{\cal M}}_2  \notag \\
\widetilde{{\cal M}}_3&=
S_{F,3}\left(\frac{\mu A_\varphi}{2}\right)^4
\int \frac{d^4 k}{(2\pi)^4} (k_1^2-m_{\chi}^2+i\epsilon)^{-1} (k_2^2-m_{\chi}^2+i\epsilon)^{-1} (k_3^2-m_{\chi}^2+i\epsilon)^{-1}\;. \label{eq:dia2}
\eal
where the loop momenta are assigned as $k_1=k-Q$, $k_2=k-2Q$, and $k_3=k$. 
The symmetric factor is given by $S_{F,3}$. 
Diagram 3 also possesses one rotational symmetry and two reflection symmetries, as Diagram 1, so the symmetric factor is $S_{F,3}=1/2$.
One can calculate the imaginary part of pinched Diagram $\widetilde{{\cal M}}_3$ by applying the cutting rules. 
There are three possibilities to cut $\widetilde{{\cal M}}_3$ as shown in Fig.~\ref{FIG:p2_pinched}. 
We note that Diagrams 3-1 and 3-2 contribute to $n_{\varphi}=1$, and Diagram 3-3 contributes to $n_{\varphi}=2$. 
The imaginary part of each diagram is calculated by applying the cutting rules as 
\bal
{\rm Im}\widetilde{\cal M}_{3\mbox{-}1}&={\rm Im}\widetilde{\cal M}_{3\mbox{-}2}
=\pi^2 S_{F,3}\left(\frac{\mu A_\varphi}{2}\right)^4
\frac{1}{16\pi^3}B_{\rm kin}(Q^2,\xi,m_{\chi}^2)
(2Q^2+2\xi-2m^2_{\chi})^{-1}\;, \notag \\
{\rm Im}\widetilde{\cal M}_{3\mbox{-}3}
&=\pi^2 S_{F,3}\left(\frac{\mu A_\varphi}{2}\right)^4
\frac{1}{16\pi^3}B_{\rm kin}(4Q^2,m^2_{\chi},m_{\chi}^2)
(-Q^2+m^2_{\chi}-\xi)^{-1}\;,
\eal
where we have used the on-shell conditions indicated by the delta functions and applied the formula Eq.~\eqref{eq:int formula}. 
Summing these results, the imaginary part of Diagram 3 is given by
\bal\label{eq:imdia2}
{\rm Im}{\cal M}_3&=\lim_{\xi\to m_{\chi}^2}\frac{\partial }{\partial \xi}
\left(2{\rm Im}\widetilde{\cal M}_{3\mbox{-}1}+{\rm Im}\widetilde{\cal M}_{3\mbox{-}3}\right) \notag \\
&=S_{F,3}\frac{m_{\varphi}^4}{16\pi}\left(\frac{\mu A_\varphi}{2m_\varphi^2}\right)^4\left(-\frac{1+\beta_1^2}{\beta_1}+\beta_{2}\right). 
\eal
%with $\beta_{N_\phi=2}=(1-m_\chi^2/m_{\varphi}^2)^{1/2}$.
In this way, from Diagram 3, we obtain the NLO contributions for the $n_{\varphi}=1$ and the LO order contributions for the $n_\varphi=2$ in the $\beta$ expansions. 

There is also Diagram 2, which is pinched into Diagram 1 as shown in Fig.~\ref{FIG:p2_pinched}.
One can compute the contribution as 
\bal
\label{eq:im_M_2}
{\rm Im}{\cal M}_2=-S_{F,2}\frac{m_{\varphi}^4}{16\pi} \left(\frac{\mu A_\varphi}{2 m_{\varphi}^2}\right)^4
\frac{1+\beta_1^2}{\beta_1^3} \;,
\eal
where two rotational symmetries and two reflection symmetries exist for this diagram, leading to $S_{F,2}=1/4$.
Diagram 2 gives the lowest-order contributions in a small $\beta$ expansion for $\theta^2$ terms because the corresponding bubble diagram involves two sets of common loop momentum as can be seen from Fig.~\ref{FIG:p=2} and they are cut. 
That is to say, two derivatives on the cut propagators arise by the pinch technique. 
Summing these results, one obtains the contributions for $\theta^4$ for the decay rate of the scalar condensate with $n_\varphi =1$:
\bal
\label{eq:nphi=1 p=2 result}
\boxed{
\Gamma_{n_\varphi=1}^{p=2}=-\frac{m_\varphi^4}{32\pi  }\left(\frac{\mu A_\varphi}{2m_\varphi^2}\right)^4\frac{1}{\beta^3_1}\Big(1+3\beta^2_1+2\beta^4_1\Big)\;.
}
\eal
One can also obtain the LO term for $n_\varphi=2$ as 
\bal
\label{eq:nphi=2 p=2 result}
\boxed{
\Gamma_{n_\varphi=2}^{p=2}=
\frac{m_{\varphi}^4}{16\pi}\left(\frac{\mu A_\varphi}{2m_\varphi^2}\right)^4\beta_{2} }\;. 
\eal

\subsection{$p=3$}
\begin{figure}[t]
  \centering
  \includegraphics[scale=0.6]{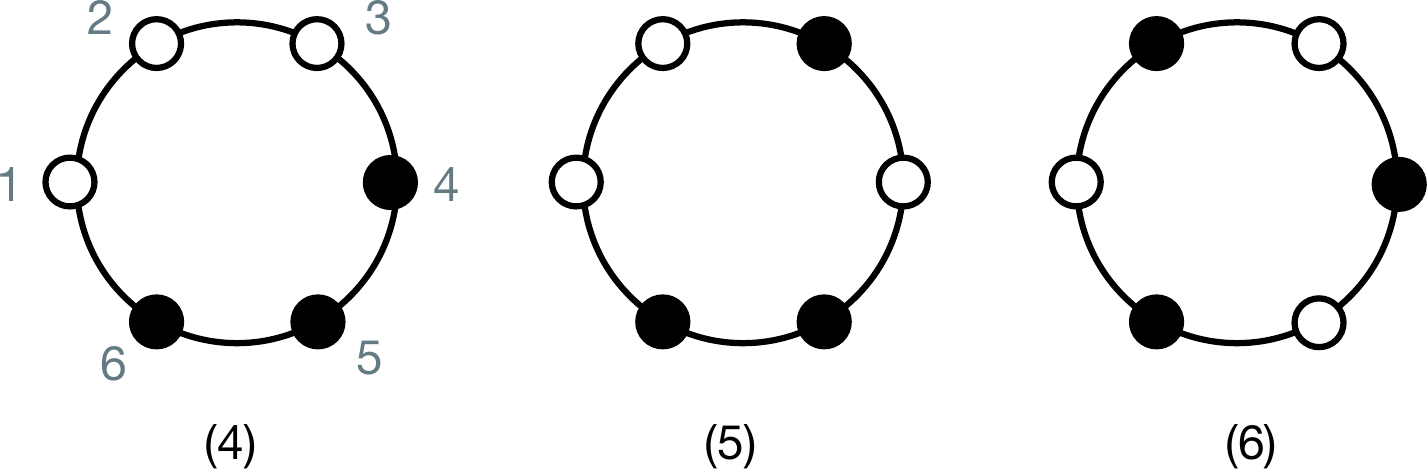}
   \caption{$p=3$ diagrams.
     }
  \label{FIG:p=3}
 \end{figure}
\begin{figure}[t]
  \centering
  \includegraphics[scale=0.6]{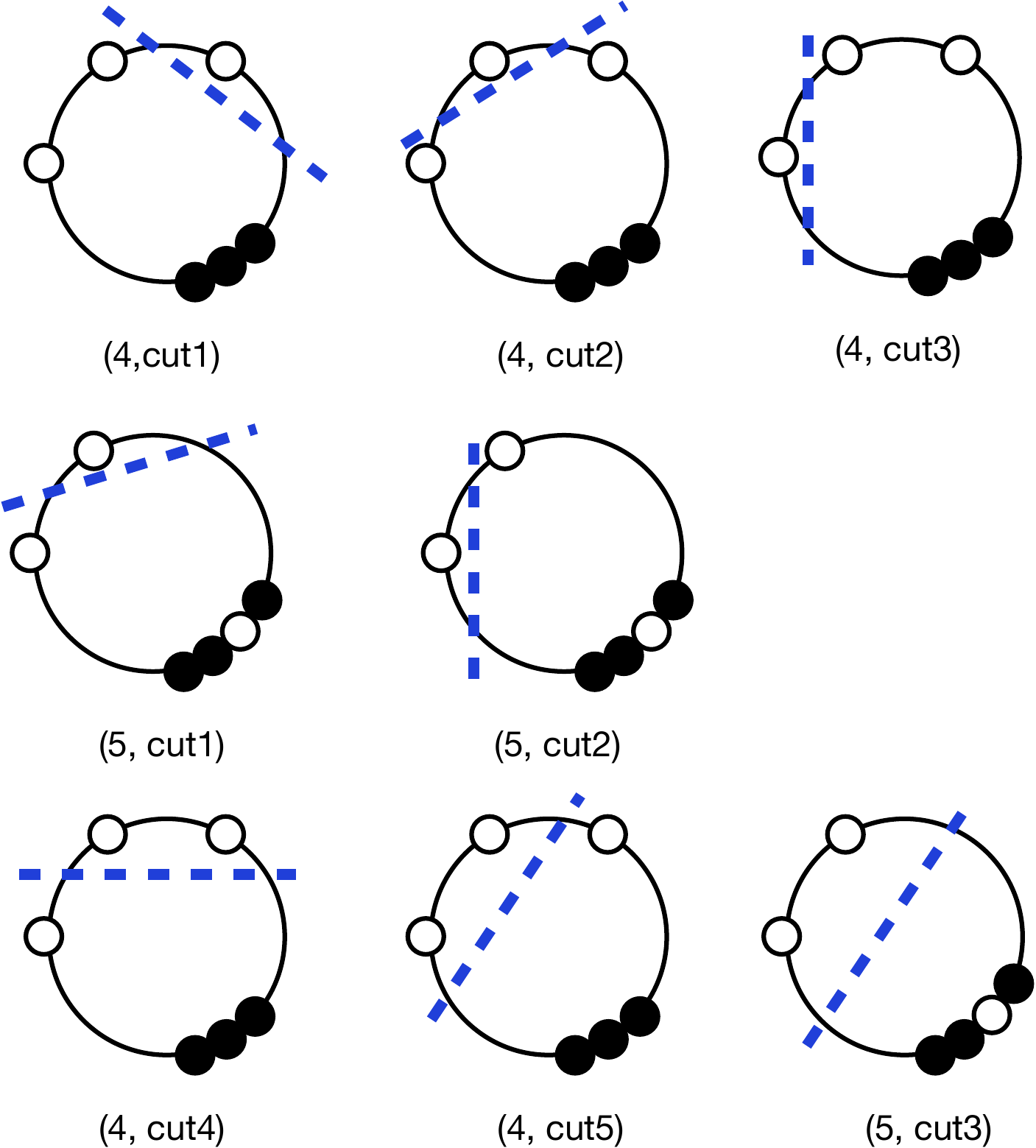}
   \caption{Possible cuts for pinched Diagrams 4 and 5, contributing $n_\varphi=1$ and $n_\varphi=2$.
     }
  \label{FIG:p=3_pinch_cut}
 \end{figure}
We then turn to the bubble diagrams with $p=3$, for which one possesses three diagrams as shown in Fig.~\ref{FIG:p=3}.\footnote{We note these are all possible diagrams that are not identical to each other under rotations, reflections, or reverse. There is a confusing diagram, such as the diagram interchanging the positions 4 and 5 in Diagram 5. We note that this diagram is generated by the same Wick contraction as Diagram 5. 
Hence, this diagram should not be counted.}. 
Depending on how the diagrams are cut as in Fig.~\ref{FIG:p=3_pinch_cut} (note that Diagram 6 is pinched to Diagram 1 and thus not shown), one obtains either the $n_\varphi=1$ contributions or the $n_\varphi=2$ contributions~\footnote{The other cut diagram exists for Diagram 4, which leads to the LO contribution for $n_\varphi=3$.}.

Focusing on the $n_\varphi=1$ contributions, the imaginary parts of the $p=3$ diagrams are written by 
%summing six cut diagrams.  
\bal
{\rm Im}{\cal M}_4^{n_\varphi=1}&=
-S_{F,4}m_\varphi^4 \left(\frac{\mu A_\varphi}{2 m_\varphi^2}\right)^6\frac{1}{1536\pi \beta_1^3} 
\left(12+4\beta_1^2+19\beta_1^4\right)\;, \notag  \\
{\rm Im}{\cal M}_5^{n_\varphi=1}&=S_{F,5}
 m_\varphi^4 \left(\frac{\mu A_\varphi}{2 m_\varphi^2}\right)^6\frac{1}{64\pi \beta_1^5} 
\left(-3+5\beta_1^4+6\beta_1^6\right)\;, \notag  \\
{\rm Im}{\cal M}_6^{n_\varphi=1}
&=-S_{F,6}m_\varphi^4 \left(\frac{\mu A_\varphi}{2 m_\varphi^2}\right)^6\frac{1}{128\pi \beta_1^7} 
\left(5+2\beta_1^2+\beta_1^4\right)\;.
\eal
With respect to the symmetric factor for these diagrams, Diagram 4 possesses $S_{F,4}=1/2$, similar to Diagrams 1 and 3. For Diagram 5, $S_{F,5}=1$ is obtained since only an identical transformation exists for rotation and reflection.
Due to the existence of three rotational symmetries and two reflection symmetries for Diagram 6, the symmetric factor becomes $S_{F,6}=1/6$. 

We note that, as mentioned above, the order of $\beta_1$ for each diagram relates to the number of $\xi$ derivatives that operate on the cut propagators in each diagram. 
For instance, Diagrams 4-1, 4-2 and 5-1 have the one, two and three $\xi$ derivatives of $B_{\rm kin}$, leading to the contributions of $\beta_1^{-1}$, $\beta_1^{-3}$ and $\beta_1^{-5}$ (and higher powers), respectively.
Note that Diagram 4 has two $\xi$ derivatives but they act on only one cut propagator in Diagram 4-1, while acting on two cut propagators in Diagram 4-2.
Summing these results, we can obtain the $n_\varphi=1$ contributions from the $p=3$ diagrams as 
\bal
\label{eq:nphi=1 p=3 result}
\boxed{
\Gamma_{n_\varphi=1}^{p=3}
=\frac{{m_\varphi^4}}{768\pi \beta_1^7}
\left(\frac{\mu A_\varphi}{2m_\varphi^2}\right)^6
\left(-60-96\beta_1^2{-24\beta_1^4}+116\beta_1^6+125\beta_1^8\right). 
}
\eal

As shown in Fig.~\ref{FIG:p=3_pinch_cut}, other possible cuts exist, which produce contributions to $n_\varphi=2$. 
%We here turn to compute the diagrams of $p=3$ contributing $N_\varphi=2$. 
%The $p=3$ diagrams are shown in Fig.~\ref{FIG:p3}.
Following the same procedure, one obtains the $n_\varphi=2$ contributions from the $p=3$ bubble diagrams as 
\bal
\label{eq:nphi=2 p=3 result}
\boxed{
\Gamma_{n_\varphi=2}^{p=3}=
{m_\varphi^4}\frac{1}{8\pi}
\left(\frac{\mu A_\varphi}{2m_\varphi^2}\right)^6
\left(\frac{1}{6\beta_2}-\frac{4}{3}\beta_2\right) \;.
}
\eal
Both Diagrams 4 and 5 have one $\xi$ derivative and thus contribute at the same order of $\beta$.

\section{Equality at lower orders} \label{sec:equality}
In Sec.~\ref{sec:diagrammatic} and Sec.~\ref{sec:parametric}, we have discussed the way of calculations in the parametric-resonance approach and Feynman-diagrammatic approach, and showed the specific analytical results derived from each method.
In this section, we demonstrate that the results obtained from these approaches are consistent with each other. 

In the parametric-resonance approach, the results at the lattice points on the $p$-$q$ plane are obtained along the diagonal directions as shown by the red, blue, and green lines in Fig.~\ref{FIG:lattice_points}. 
For the case of $n_\varphi=1\ (2)$, we have obtained results up to the NNLO (NLO) terms in the $\beta$ expansion for arbitrary orders of $\theta$. 
On the other hand, in the Feynman-diagrammatic approach, the results are obtained along the vertical direction of the $q$-$p$ plane. 
Thus, for a specific order of $\theta$, results for all orders in the $\beta$ expansion can be obtained.
Computing all the bubble diagrams with $p\lesssim3$, we have obtained all the results in the case of $n_\varphi=1$ and $n_\varphi=2$. 

We have indeed obtained the same results using these two methods. 
As a concrete example, we focus on the LO results obtained from the parametric-resonance approaches, i.e., Eqs.~\eqref{eq:nphi=1 LO result} and \eqref{eq:Gam_nphi=2_para}. 
Expanding them with respect to $p$ yields 
for $n_\varphi=1$
\bal
\Gamma_{n_{\varphi}=1}^{\rm LO}&=\frac{m_\varphi^4\beta_1}{16\pi}\left(\frac{\mu A_\varphi}{2m_\varphi^2}\right)^2-\frac{m_\varphi^4}{32\pi\beta_1^3}\left(\frac{\mu A_\varphi}{2m_\varphi^2}\right)^4-\frac{5m_\varphi^4}{64\pi \beta_1^7}\left(\frac{\mu A_\varphi}{2m_\varphi^2}\right)^6+\dots
\\
\Gamma_{n_{\varphi}=1}^{\rm NLO}&=\qquad \qquad \qquad \quad 
-3\frac{m_\varphi^4}{32\pi\beta_1}\left(\frac{\mu A_\varphi}{2m_\varphi^2}\right)^4-\frac{m_\varphi^4}{8\pi \beta_1^5}\left(\frac{\mu A_\varphi}{2m_\varphi^2}\right)^6+\dots
\\
\Gamma_{n_{\varphi}=1}^{\rm NNLO}&=\qquad \qquad \qquad \quad 
-\frac{m_\varphi^4\beta_1}{16\pi}\left(\frac{\mu A_\varphi}{2m_\varphi^2}\right)^4-\frac{m_\varphi^4}{32\pi \beta_1^3}\left(\frac{\mu A_\varphi}{2m_\varphi^2}\right)^6+\dots
\eal
for $n_\varphi=2$
\bal
\Gamma_{n_{\varphi}=2}^{\rm LO}&=\frac{m_\varphi^4\beta_2}{16\pi}\left(\frac{\mu A_\varphi}{2m_\varphi^2}\right)^4+\frac{m_\varphi^4}{48\pi\beta_2}\left(\frac{\mu A_\varphi}{2m_\varphi^2}\right)^6+\dots \\
%-\frac{25m_\varphi^4}{4608\pi \beta_2^3}\left(\frac{\mu A_\varphi}{2m_\varphi^2}\right)^8+\dots
\Gamma_{n_{\varphi}=2}^{\rm NLO}&=\quad\quad\quad\quad\quad\quad\quad
-\frac{m_\varphi^4\beta_2}{6\pi}\left(\frac{\mu A_\varphi}{2m_\varphi^2}\right)^6+\dots
\eal
One can see that $\Gamma_{n_\varphi=1}$ above is consistent with  Eqs.~\eqref{eq:nphi=1 p=1 result}, \eqref{eq:nphi=1 p=2 result} and the first three terms of \eqref{eq:nphi=1 p=3 result}. 
In addition, $\Gamma_{n_\varphi=2}$ above is consistent with  Eq.~\eqref{eq:nphi=2 p=2 result} and Eq.~\eqref{eq:nphi=2 p=3 result}.

Actually, it is not difficult to reproduce $\Gamma_{n_{\varphi}=1}^{\rm LO}$ at all orders in the Feynman-diagrammatic approach.
This follows from the observation that the largest number of $\xi$ acting on the cut propagators arises from the diagrams with pair-wise insertions of $\varphi_+$ and $\varphi_-$ (namely, black and white circles appear one after the other, e.g., Diagrams 2 and 6 or Fig.~3 of Ref.~\cite{Matsumoto:2007rd}).
Such diagrams are pinched into Diagram 1 and the contribution for $\theta^{2p}$ term is given by
\bal
\lim_{\xi_1\to m_\chi^2} \lim_{\xi_2\to m_\chi^2} \frac{2}{2p(p-1)!(p-1)!} \frac{\partial^{p-1}}{\partial \xi_1^{p-1}}
\frac{\partial^{p-1}}{\partial \xi_2^{p-1}} 
2\pi^2 \left(\frac{\mu A_\varphi}{2}\right)^{2p}
\frac{1}{32\pi^3}B_{\rm kin}(Q^2,\xi_1,\xi_2)\;.
\eal
Here $2/(2p)$ is the ratio of the symmetry factors for $p=1$ and $p$ ($p$ rotations and $2$ reflections).
By changing the variable $\xi_1 = m_\chi^2 + \eta_1$ and $\xi_2 = m_\chi^2 + \eta_2$, one needs to evaluate
\bal
 {\rm Im}{\cal M}_p^{\rm pair} =
 \lim_{\eta_1\to 0} \lim_{\eta_2\to 0} \frac{\partial^{p-1}}{\partial \eta_1^{p-1}}
\frac{\partial^{p-1}}{\partial \eta_2^{p-1}} \sqrt{\beta_1^2-\frac{2(\eta_1+\eta_2)}{m_\varphi^2} + \frac{(\eta_1-\eta_2)^2}{m_\varphi^4}} \;.
\eal
Noting that $(\eta_1-\eta_2)^2$ can be dropped at the leading order of $\beta_1$, one can easily see that Eqs.~\eqref{eq:nphi=1 LO result} is reproduced.

\section{Conclusion and discussion} \label{sec:conclusion}

We have studied quantum field theoretic description of energy dissipation (or decay) of scalar condensate via interaction with daughter particles.
Two approaches can be found in the literature: one is based on parametric resonance in mode functions of a daughter particle~\cite{Yoshimura:1995gc}; the other is based on $S$-matrix of a coherent state and Feynman-diagrammatic perturbation~\cite{Matsumoto:2007rd}.
We have modified the latter in a way that manifests what we are computing and does not include the unwanted Feynman diagrams.
Though they look different both conceptually and computationally, we point out that they are computing the same quantity: vacuum-to-vacuum transition of the daughter particle in the presence of background scalar condensate.

To be concrete, we have considered the narrow resonance (or small amplitude) regime.
The apparent difference between these two approaches is manifest in the double expansion of decay rate in terms of the amplitude and velocity (or mass difference between the mother particle $\varphi$ and daughter particle $\chi$).
The parametric-resonance approach provides a certain combination of orders of the amplitude and velocity at once.
On the other hand, the Feynman-diagrammatic approach provides all orders of the velocity for a given order of the amplitude.
By explicitly computing the lower order contributions in both the two approaches, we have demonstrated their equivalence.

{
On the other hand, the following question is to be more closely examined: whether the two approaches give the same result for any time interval $T$ or not.
In the parametric-resonance approach, Eq.~\eqref{eq:vac-to-vac} is valid for any $T$, but Eq.~\eqref{eq:Gamma_wrt_k} is valid only for a sufficiently large $T$ so that the decay of the vacuum-to-vacuum transition amplitude is predominantly determined by the growing mode and approximated by the growth factor.
Therefore, it is not clear how the decay exponent of the transition amplitude behaves for a small $T$. (In general, it is not necessarily proportional to $T$.)
In the Feynman-diagrammatic approach, Eq.~\eqref{eq:Gamma_wrt_S} would be fully non-perturbative, but actual computation is performed perturbatively in terms of small amplitude and the validity range of $T$ is not quite clear. (It should be larger than the oscillation frequency of scalar condensate so that energy conservation is well satisfied.)
Interestingly, the contribution of each Feynman diagram to the decay exponent is proportional to $T$, coming from the delta function for the energy-momentum conservation.

Probably related to the validity range of $T$, we remark that the differential decay rate in the Feynman-diagrammatic approach [see the discussion below Eq.~\eqref{eq:int formula}] does not coincide with the growth factor.
Concerning the decay process ($n_\varphi  = 1$) at the leading order, the former is proportional to the amplitude squared and the momentum distribution of $\delta(k - n_\varphi m_\varphi \beta_{n_\varphi} / 2)$, while the latter is proportional to the amplitude and its momentum distribution has the width proportional to the amplitude.
Actually, this makes our result (equivalence) more non-trivial.

To answer the question about $T$ and understand the equivalence better, the followings are worth investigating.
One is to note that the decay exponent at any $T$ can be evaluated in terms of the distribution function of a daughter particle [see Eq.~\eqref{eq:phase_space}], which can be obtained numerically (probably also analytically in a perturbation theory).
Another is to study the fermionic daughter particle, which does not allow an exponential growth of the mode function because of Pauli blocking~\cite{Baacke:1998di, Greene:1998nh, Greene:2000ew}.
We would expect that there is no difference between the boson and fermion for a small $T$, but Bose enhancement and Pauli blocking makes a difference for a large $T$~\cite{Asaka:2010kv}.
%Ref.~\cite{Li:2025wwl} solves the Boltzmann equation with a finite-width stimulated production, and shows that the result is in a fair agreement with that of the Bogoliubov-transformation approach (see, e.g., Ref.~\cite{Kofman:1997yn}), which directly relates the growing mode and the momentum distribution of the daughter particles.
}

In addition, the followings would be worth investigating in the future.
%We have focused on a scalar $\chi$ with a cubic interaction $\varphi \chi^{2}$ by following the previous literature~\cite{Yoshimura:1995gc, Matsumoto:2007rd}.
%A fermionic $\chi$ requires an extension of the formulation in Sec.~\ref{sec:review parametric} or Ref.~\cite{Yoshimura:1995gc} to a fermionic harmonic oscillator, though Pauli-Blocking suppresses the production of daughter particles {(which is investigated in the Bogoliubov-transformation approach; see, e.g., Ref.~\cite{Baacke:1998di})}.
We have ignored fluctuations of $\varphi$ by following the previous literature~\cite{Yoshimura:1995gc, Matsumoto:2007rd}.
Inclusion of kinetic term of $\varphi$ will be straightforward, but not inclusion of interaction between $\varphi$ and $\chi$ in the parametric-resonance approach.
We have focused on the small amplitude (or narrow resonance) regime, where the Feynman-diagrammatic expansion is justified.
On the other hand, in the large amplitude (or broad resonance) regime, one cannot rely on perturbation theory and thus needs to employ, say, a dressed-propagator (and its Schwinger-time expression) approach~\cite{Taya:2022gzp, Xu:2023rqf}.

\appendix
\section{Apparent divergence in the imaginary part of the bubble diagrams} \label{sec:apparent divergence}
We would like to discuss the subtleness in applying the cutting rule to the bubble diagrams. 
Due to the $\varphi$ insertion the momentum of $\varphi$, $Q=(m_\varphi,{\bf 0})$ flows in $\chi$ propagators. 
Since the incoming $\varphi$ and outgoing $\varphi$ necessary appear in pair (e.g., see the concrete diagrams with specified $p$, Figs.~\ref{FIG:p=1}, \ref{FIG:p=2} and ~\ref{FIG:p=3}), some of $\chi$ propagators have a common momentum. 
In this case, evaluations of diagrams after applying cut are indeterminate because, depending on cut diagrams, on-shell conditions $k_i^2=m_\chi^2$ are imposed for the propagator $(k_i^2-m_\chi^2+i\epsilon)^{-1}$, which leads to apparent divergence. 
To obtain finite results, one needs to regulate it. 
One of the ways to treat this apparent divergence is performed in Ref.~\cite{Matsumoto:2007rd}. In this earlier work, in applying the cutting rules to the bubble diagrams, the mass of all propagators is separately replaced by dimension 2 parameter $\xi_i$, for which the limit $\xi_i\to m_\chi^2$ is taken afterward. 
In this appendix, let us discuss how such an apparent divergence arises and cancels with each other. 

To be concrete, we focus on Diagram 2 in the Fig.~\ref{FIG:p=2}. 
As discussed in the main text, the amplitude is given by
\bal
i{{\cal M}}_2&=
\frac{1}{4}\left(\frac{\mu A_\varphi}{2}\right)^4
\int \frac{d^4 k}{(2\pi)^4} (k_1^2-m_{\chi}^2+i\epsilon)^{-1} (k_2^2-m_{\chi}^2+i\epsilon)^{-1} \notag \\ \;
&\times (k_3^2-m_{\chi}^2+i\epsilon)^{-1}(k_4^2-m_{\chi}^2+i\epsilon)^{-1}. 
\eal
where loop momenta are assigned by $k_1=k_3=k+Q$ and $k_2=k_4=k$. 
To evade divergence caused by applying cut, we replace the mass of $\chi$ by $\xi$ parameter as
\bal
i{{\cal M}}_2&=\lim_{\xi_1\to m_\chi^2}\dots\lim_{\xi_4\to m_\chi^2}
\frac{1}{4}\left(\frac{\mu A_\varphi}{2}\right)^4
\int \frac{d^4 k}{(2\pi)^4} \notag \\
&\times 
(k_1^2-\xi_1+i\epsilon)^{-1} (k_2^2-\xi_2+i\epsilon)^{-1}
(k_3^2-\xi_3+i\epsilon)^{-1}(k_4^2-\xi_4+i\epsilon)^{-1} \;. 
\eal
We then apply the cutting rule to evaluate the imaginary part of ${\cal M}_2$.
For Diagram 2, there are four possible cuts:
\bal
{\rm Im}{{\cal M}}_2={\rm Im} \left({\cal M}_2^{{\rm cut} 12}+ {\cal M}_2^{{\rm cut} 34}+{\cal M}_2^{{\rm cut} 23}+{\cal M}_2^{{\rm cut} 14}\right)
\eal
where for the cut ${\cal M}^{{\rm cut}ij}_2$ we have applied the cut for the propagators with the momenta $k_i$ and $k_j$. 
Each cut is given by
\bal
 {\rm Im}{\cal M}^{{\rm cut}ij}_2=\frac{1}{4}\left(\frac{\mu A_\varphi}{2}\right)^4
 \lim_{\xi_1\to m_\chi^2}\dots\lim_{\xi_4\to m_\chi^2}
 \frac{c_{ij}}{16\pi}\frac{B_{\rm kin}^{ij}}{(\xi_1-\xi_3)(\xi_2-\xi_4)},
\eal
where $c_{ij}=+1$ for $(i,j)=(1,2),\ (3,4)$, $c_{ij}=-1$ for $(i,j)=(2,3),\ (1,4)$, and $c_{ij}=0$ for others. 
We have also introduced the short-handed notation for the kinematical function $B_{\rm kin}^{ij}\equiv B_{\rm kin} (Q^2,\xi_i,\xi_j)$. 
To be clear the structure of the divergent part for each cut diagram,  
we expand $\xi_1$ and $\xi_2$ by introducing $\eta_i$ $(i=1,2)$ as $\xi_1= \xi_3+\eta_1$ and  $\xi_2= \xi_4+\eta_2$. 
The cut $ {\rm Im}{\cal M}^{{\rm cut}ij}_2$ is then reduced as
\bal
\label{eq:cut12}
{\rm Im} {\cal M}_2^{{\rm cut} 12}&= 
\frac{1}{4}\left(\frac{\mu A_\varphi}{2}\right)^4\lim_{\substack{\xi_3\to m_\chi^2 \\ \xi_4\to m_\chi^2}}\lim_{\substack{\eta_1\to 0 \\ \eta_2\to 0}}\frac{1}{16\pi}\frac{1}{\eta_1 \eta_2}
\Bigg[B_{\rm kin}^{34}+\eta_1\frac{\partial B_{\rm kin}^{14}}{\partial \xi_1}+\eta_2\frac{\partial B_{\rm kin}^{32}}{\partial \xi_2} \notag \\
&+\eta_1\eta_2\frac{\partial^2 B_{\rm kin}^{12}}{\partial \xi_1 \partial \xi_2}
+\frac{\eta_1^2}{2}\frac{\partial^2 B_{\rm kin}^{14}}{\partial \xi_1^2}
+\frac{\eta_2^2}{2}\frac{\partial^2 B_{\rm kin}^{32}}{\partial \xi_2^2}+{\cal O}(\eta_1^a \eta_2^b)
\Bigg]_{\substack{\xi_1=\xi_3\\ \xi_2=\xi_4}}, \\
%----------
{\rm Im} {\cal M}_2^{{\rm cut} 34}&=\frac{1}{4}\left(\frac{\mu A_\varphi}{2}\right)^4\lim_{\substack{\xi_3\to m_\chi^2 \\ \xi_4\to m_\chi^2}}\lim_{\substack{\eta_1\to 0 \\ \eta_2\to 0}}\frac{1}{16\pi}\frac{1}{\eta_1 \eta_2} B_{\rm kin}^{34}, 
 \\
%----------
{\rm Im} {\cal M}_2^{{\rm cut} 14}&=-\frac{1}{4}\left(\frac{\mu A_\varphi}{2}\right)^4\lim_{\substack{\xi_3\to m_\chi^2 \\ \xi_4\to m_\chi^2}}\lim_{\substack{\eta_1\to 0 \\ \eta_2\to 0}}
\frac{1}{16\pi}\frac{1}{\eta_1 \eta_2}
\Bigg[B_{\rm kin}^{34}+\eta_1\frac{\partial B_{\rm kin}^{14}}{\partial \xi_1} 
+\frac{\eta_1^2}{2}\frac{\partial^2 B_{\rm kin}^{14}}{\partial \xi_1^2}+{\cal O}(\eta^3_1)
\Bigg]_{\substack{\xi_1=\xi_3}},
 \\
%----------
{\rm Im} {\cal M}_2^{{\rm cut} 23}&=-\frac{1}{4}\left(\frac{\mu A_\varphi}{2}\right)^4\lim_{\substack{\xi_3\to m_\chi^2 \\ \xi_4\to m_\chi^2}}\lim_{\substack{\eta_1\to 0 \\ \eta_2\to 0}}
\frac{1}{16\pi}\frac{1}{\eta_1 \eta_2}
\Bigg[B_{\rm kin}^{34}+\eta_2\frac{\partial B_{\rm kin}^{32}}{\partial \xi_2} 
+\frac{\eta_2^2}{2}\frac{\partial^2 B_{\rm kin}^{32}}{\partial \xi_2^2}
+{\cal O}(\eta^3_2) \Bigg]_{\xi_2=\xi_4},
\eal
with $a,b$ being a integer satisfying $a+b\geq 3$. We note that, except for the fourth term ($\eta_1 \eta_2 \partial^2 B_{\rm kin}^2/\partial \xi_1 \partial \xi_2$) in Eq.~\eqref{eq:cut12}, all the terms are divergent or indeterminate in the limits $\xi_i\to m_\chi^2$ and $\eta_i\to0$ $(i=1,2)$. 
The higher order terms ${\cal O}(\eta_1^a \eta_2^b)$ ($a+b\geq 3$) converge to zero. 
One can easily see that by adding all cut diagrams, such divergent or indeterminate terms cancel out with each other. 
As a result, the imaginary part of the ${\cal M}_2$ is given by the finite part as
\bal
{\rm Im} {\cal M}_2^{ }=\frac{1}{4}\left(\frac{\mu A_\varphi}{2}\right)^4
\lim_{\substack{\xi_3\to m_\chi^2 \\ \xi_4\to m_\chi^2}}\frac{1}{16\pi}\frac{\partial^2 B_{\rm kin}^{34}}{\partial \xi_3 \partial \xi_4},
\eal
which is consistent with the result in the main text, i.e., Eq.~\eqref{eq:im_M_2}. 
% By this replacement, the divergence part is regularized. 
% Although each cut diagram contains the divergence, one obtains finite results after summing all possible cuts. 

The pinch technique used in this study corresponds to another type of regularization for the concerned divergence. 
The essential point for this procedure is that we apply the pinch technique before applying the cutting rules. 
The pinch technique removes the common propagator, so that the pinched amplitude involves a set of propagators only with the different momentum. 
Hence, no issue of divergence arises in the application of the cutting rules.  
That is to say, one can directly calculate the finite part of the imaginary part of the bubble diagrams. 
%We have checked that, for the dominant contributions in $p=2$ diagrams 2 in Fig.~\ref{FIG:p=2}, the results derived in our procedure coincide with those of Ref.~\cite{Matsumoto:2007rd}. 
For Diagram 3 in Fig.~\ref{FIG:p=2}, we have calculated the imaginary part of the diagram by using the Passarino-Veltman function, the $D_0$ function~\cite{Passarino:1978jh}. 
By using {\tt Looptools}~\cite{Hahn:1998yk}, we have numerically (by introducing a tiny adiabatic parameter $\epsilon$) checked that this is consistent with the results obtained from our procedure.

\section{Oscillation average} \label{sec:oscillation average}
In this appendix, we discuss the calculation of the scalar condensate decay using the oscillation average. 
This provides the most naive approach to evaluating the decay rate of the scalar condensate. 

The simplest expression for the scalar decay rate per unit volume can be written as $m_\varphi\Gamma_\varphi=\gamma_\varphi \rho_\varphi$, where $\gamma_\varphi$ corresponds to the decay rate of single scalar particle in the vacuum, $\gamma_\varphi= \mu^2\beta_1/(32\pi m_\varphi)$ and $\rho_\varphi$ denotes the energy density of the single scalar particle. 
It should be noted that this describes the decay of the one-particle state $\varphi$ in the vacuum rather than $\varphi$ in the oscillation. 
On the other hand, in Ref.~\cite{Ahmed:2021fvt}, the influence of the oscillation of $\varphi$ is taken into account by incorporating the oscillation effect for the mass of the daughter particle. Thus,
\bal
m_\varphi\Gamma_\varphi=\braket{\left.\gamma_\varphi\right|_{\beta_1  \rightarrow \beta_1^{\rm eff}} }_{\rm osc.} \rho_\varphi \;,
\eal
where the kinematical function $\beta_1$ is replaced by $\beta_1^{\rm eff}=\sqrt{1-4m_{\chi{\rm eff}}^2/m_\varphi^2}$ with $m_{\chi{\rm eff}}^2=m_\chi^2+\mu A_\varphi \cos(m_\varphi t)$.
The bracket denotes the oscillation average: $\braket{Q}_{\rm osc.}= \left( \int_{0}^{T} dt Q \right)/ T$ with oscillation period $T$.
By computing this expression, we obtain
\bal \label{eq:naive_oscillation}
\Gamma_\varphi=-\frac{m_\varphi^4}{64\pi}\sum_{p=1}^\infty\beta^{5-4p}_1\left(\frac{\mu A_\varphi}{m_\varphi^2}\right)^{2p}
\frac{(4p-7)!!}{\{(p-1)!\}^2} \;.
\eal
This looks similar to the LO result in the parametric-resonance approach, i.e., Eq.~\eqref{eq:nphi=1 LO result}, but is different by a factor of $ 2^{2(p-1)} p$~\footnote{We note that the $p=1$ contribution in Eqs.~\eqref{eq:nphi=1 LO result} [as well as \eqref{eq:nphi=1 p=1 result}] and \eqref{eq:naive_oscillation}, reproduce $m_\varphi\Gamma_\varphi=\gamma_\varphi \rho_\varphi$ with $\gamma_\varphi$ being the vacuum value. While Eq.~\eqref{eq:naive_oscillation}, after multiplying the appropriate factor [Eq.~\eqref{eq:improved_oscillation}], coincides with Eq.~\eqref{eq:nphi=1 LO result} for any $p$, calculations in Refs.~\cite{Ichikawa:2008ne,Garcia:2020eof,Garcia:2020wiy,Clery:2021bwz,Clery:2022wib} provides only $\gamma_{\varphi}$ but not $p \geq 2$ nor NLOs.}.

We further incorporate the observation that decay is associated with only kinetic energy (not potential energy)~\cite{Kolb:1990vq} by
\bal
\label{eq:improved_oscillation}
m_\varphi\Gamma_{\varphi}=\Braket{\left.\gamma_{\varphi}\right|_{ \beta_1  \rightarrow \beta_1^{\rm eff} }  \left(\frac{d \varphi}{dt} \right)^{2}}_{\rm osc.}\;.
\eal
Note that $\braket{(d\varphi/dt)^2}_{\rm osc.} = \rho_\varphi$ for a coherently oscillating scalar in a quadratic potential.
%By computing this expression, we obtain
%\bal
%\Gamma_\varphi=\frac{m_\varphi^4 \beta_1^{5-4p}}{64\pi}
%\left(\frac{\mu A_\varphi}{2m_\varphi^2}\right)^{2p}\frac{2^{5-4p}}{5-4p}
%\frac{(4p-4)!}{\{(2p-2)!\}^2}\frac{(2p-3)!}{p!(p-2)!} \;.
%\eal
By computing this expression, we find that this reproduces the LO (but not higher order) result in the parametric-resonance approach, i.e., Eq.~\eqref{eq:nphi=1 LO result}. 

\acknowledgments
A. K. thanks Shigeki Matsumoto and Tomo Takahashi for useful discussions.
A. K. acknowledges partial support from Norwegian Financial Mechanism for years 2014-2021, grant nr 2019/34/H/ST2/00707; and from National Science Centre, Poland, grant DEC-2018/31/B/ST2/02283.
K.S. acknowledges support by Japan Society for the Promotion of Science (JSPS) under grant 25K17379 and 21K20363.

\bibliography{biblio} 
\bibliographystyle{JHEP}

% \begin{thebibliography}{99}
% \end{thebibliography}

\end{document}